\documentclass[fleqn,usenatbib]{mnras}

\usepackage[T1]{fontenc}
\usepackage{ae,aecompl,threeparttable}
\usepackage{caption}

\usepackage{graphicx,psfig,epsfig,epsf}	
\usepackage{amsmath}	
\usepackage{amssymb}	

\newcommand{\U}[3]{$#1_{-#3}^{+#2}$}

\title[Wind-instability outbursts in HLX-1]{Outbursts of the intermediate-mass black hole HLX-1:\\ a wind instability scenario}

\author[R. Soria et al.]{Roberto Soria,$^{1,2,3}$\thanks{E-mail: roberto.soria@curtin.edu.au (RS)}
Aina Musaeva,$^{2}$
Kinwah Wu,$^{4}$
Luca Zampieri,$^{5}$
Sara Federle,$^{5}$
\newauthor
Ryan Urquhart,$^{1}$
Edwin van der Helm,$^{6}$
and Sean Farrell,$^{2}$
\vspace{0.15cm}
\\
$^{1}$International Centre for Radio Astronomy Research, Curtin University, GPO Box U1987, Perth, WA 6845, Australia\\
$^{2}$Sydney Institute for Astronomy, School of Physics A28, The University of Sydney, Sydney NSW 2006, Australia\\
$^{3}$National Astronomical Observatories, Chinese Academy of Sciences, Beijing 100012, China\\
$^{4}$Mullard Space Science Laboratory, University College London, Holmbury St.~Mary, Dorking, Surrey, RH5 6NT, UK\\
$^{5}$INAF, Astronomical Observatory of Padova, vicolo dell'Osservatorio 5, I-35122 Padova, Italy\\
$^{6}$Leiden Observatory, Leiden University, PO Box 9513, NL-2300 RA, Leiden, the Netherlands
}

\date{Accepted 2017 April 08. Received 2017 March 31; in original form 2017 November 15}

\pubyear{2017}

\begin{document}
\label{firstpage}
\pagerange{\pageref{firstpage}--\pageref{lastpage}}
\maketitle

\begin{abstract}
We model the intermediate-mass black hole HLX-1, using the {\it Hubble Space Telescope}, {\it XMM-Newton} and {\it Swift}. We quantify the relative contributions of a bluer component, function of X-ray irradiation, and a redder component, constant and likely coming from an old stellar population. We estimate a black hole mass $\approx (2^{+2}_{-1}) \times 10^4 M_{\odot}$, a spin parameter $a/M \approx 0.9$ for moderately face-on view, and a peak outburst luminosity $\approx 0.3$ times the Eddington luminosity. We discuss the discrepancy between the characteristic sizes inferred from the short X-ray timescale ($R \sim$ a few $10^{11}$ cm) and from the optical emitter ($R \sqrt{\cos \theta} \approx 2.2 \times 10^{13}$ cm). One possibility is that the optical emitter is a circumbinary disk; however, we disfavour this scenario because it would require a very small donor star. A more plausible scenario is that the disk is large but only the inner annuli are involved in the X-ray outburst. We propose that the recurrent outbursts are caused by an accretion-rate oscillation driven by wind instability in the inner disk. We argue that the system has a long-term-average accretion rate of a few percent Eddington, just below the upper limit of the low/hard state; a wind-driven oscillation can trigger transitions to the high/soft state, with a recurrence period $\sim$1 year (much longer than the binary period, which we estimate as $\sim$10 days). The oscillation that dominated the system in the last decade is now damped such that the accretion rate no longer reaches the level required to trigger a transition. Finally, we highlight similarities between disk winds in HLX-1 and in the Galactic black hole V404 Cyg.
\end{abstract}

\begin{keywords}
black hole physics -- X-rays: binaries -- X-rays: individual: HLX-1
\end{keywords}



\section{Introduction}

The transient X-ray source HLX-1 \citep{farrell09} is arguably the most convincing off-nuclear intermediate-mass black hole (IMBH) candidate. The X-ray source is unambiguously associated with an H$\alpha$-emitting optical counterpart, providing a recession speed of $\approx$7,100 km s$^{-1}$ \citep{soria13b,wiersema10}; this is consistent with the recession velocity of the surrounding galaxy cluster Abell 2877 \citep{malumuth92}. HLX-1 is projected in front of the halo of the S0 galaxy ESO\,243-49 (a member of Abell 2877), $\approx$7\arcsec from its nucleus (Figure 1), which has a systemic recession speed of $\approx$6,700 km s$^{-1}$. For this reason, we assume the same luminosity distance $\approx$92 Mpc for both ESO\,243-49 and HLX-1 (NASA/IPAC Extragalactic Database). It is unclear whether HLX-1 is orbiting around ESO\,243-49 or it is simply a chance association within Abell 2877; in either case, HLX-1 is not a foreground Galactic object. If we assume isotropic emission, the peak X-ray luminosity in each of its six fully-monitored outbursts is $L_{\rm X} \approx (1$--$1.5) \times 10^{42}$ erg s$^{-1}$ \citep{yan15}, corresponding to the Eddington luminosity of a $10^4 M_{\odot}$ BH. At outburst peak, the X-ray spectrum is well fitted with a radiatively efficient, multicolour disk model \citep{davis11,godet12,farrell12}, typical of accreting BHs in the sub-Eddington high/soft state. The peak colour temperature ($kT_{\rm in} \approx 0.25$ keV) and the inner radius of the disk ($R_{\rm in} \sqrt{\cos \theta} \sim 5$--$10 \times 10^4$ km) are self-consistent with a $\sim$10$^4 M_{\odot}$ BH accreting just below its Eddington limit \citep{godet12}. The low rms variability is also consistent with an accretion disk in the high/soft state \citep{servillat11}. The outburst evolution after the peak is similar to that of sub-Eddington Galactic BHs: in the first few weeks, a power-law tail appears, strengthens and becomes flatter, while the disk component evolves to lower luminosities and temperatures along the characteristic $L \propto T_{\rm in}^4$ track \citep{servillat11,godet12}. In each of its six fully-recorded outbursts (Figure 2), after $\sim$100--150 days and after it declined to a luminosity $\la 10^{41}$ erg s$^{-1}$, HLX-1 underwent a transition to the low/hard state, with a spectrum dominated by a power-law of photon index $\Gamma = 1.6 \pm 0.3$ \citep{servillat11,yan15}. The detection of transient radio emission during an outburst \citep{webb12} and the presence of a jet in the low/hard state \citep{cseh15} also support the interpretation that HLX-1 cycles through the canonical states of an accreting BH, with radio flaring generally observed between the hard outburst rise and the transition to the thermal dominant state \citep{fender04}.

Despite the success of this IMBH model based on canonical sub-Eddington accretion states, several questions remain unanswered; failure to solve such problems might even lead to a rejection of the canonical IMBH model, in favour for example of a beamed, highly super-Eddington stellar-mass BH \citep{lasota15}. The main outstanding problems are:
\begin{itemize}
\item what component of the optical/UV luminosity comes from the irradiated accretion disk (variable component) and what from a surrounding star cluster (constant component)? 
\item what is the size of the accretion disk and of the semi-major axis of the binary system? More specifically, can the accretion disk be small enough to explain the short outburst timescale but at the same time large enough to produce the observed optical luminosity?
\item what causes the repeated outburst behaviour and determines the recurrence timescale: thermal-viscous instability, periastron passages, other types of mass transfer instabilities, oscillations induced by radiation pressure or outflows?
\end{itemize}
In this paper, we will discuss those sets of questions, re-examine published and unpublished observational results, and propose a new scenario that may be consistent with all the data. We will determine the BH mass and other system parameters consistent with this scenario.

\begin{figure}
\hspace{-0.3cm}
\psfig{figure=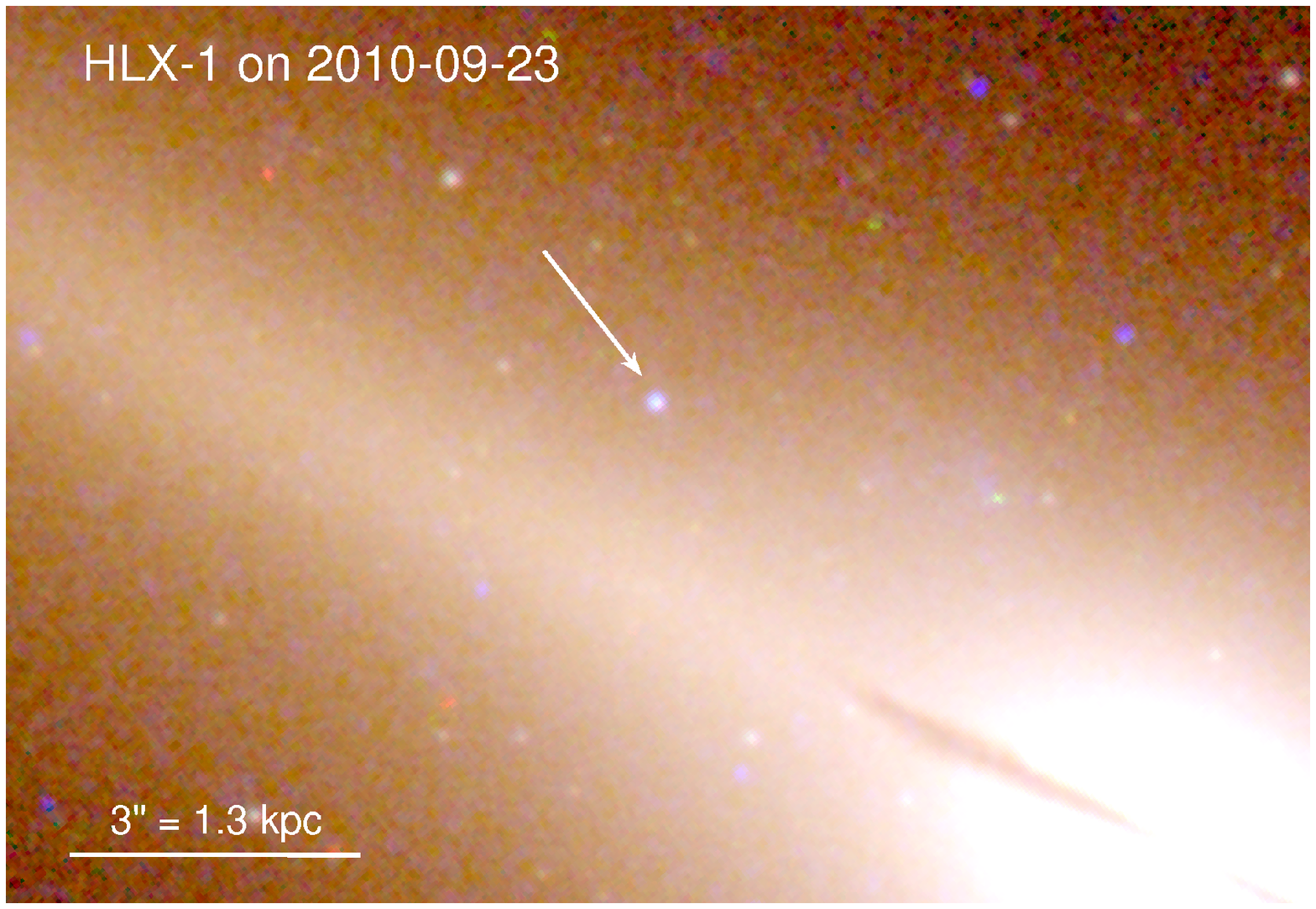,width=8.8cm,angle=0}\\

\vspace{-0.5cm}
\hspace{-0.3cm}
\psfig{figure=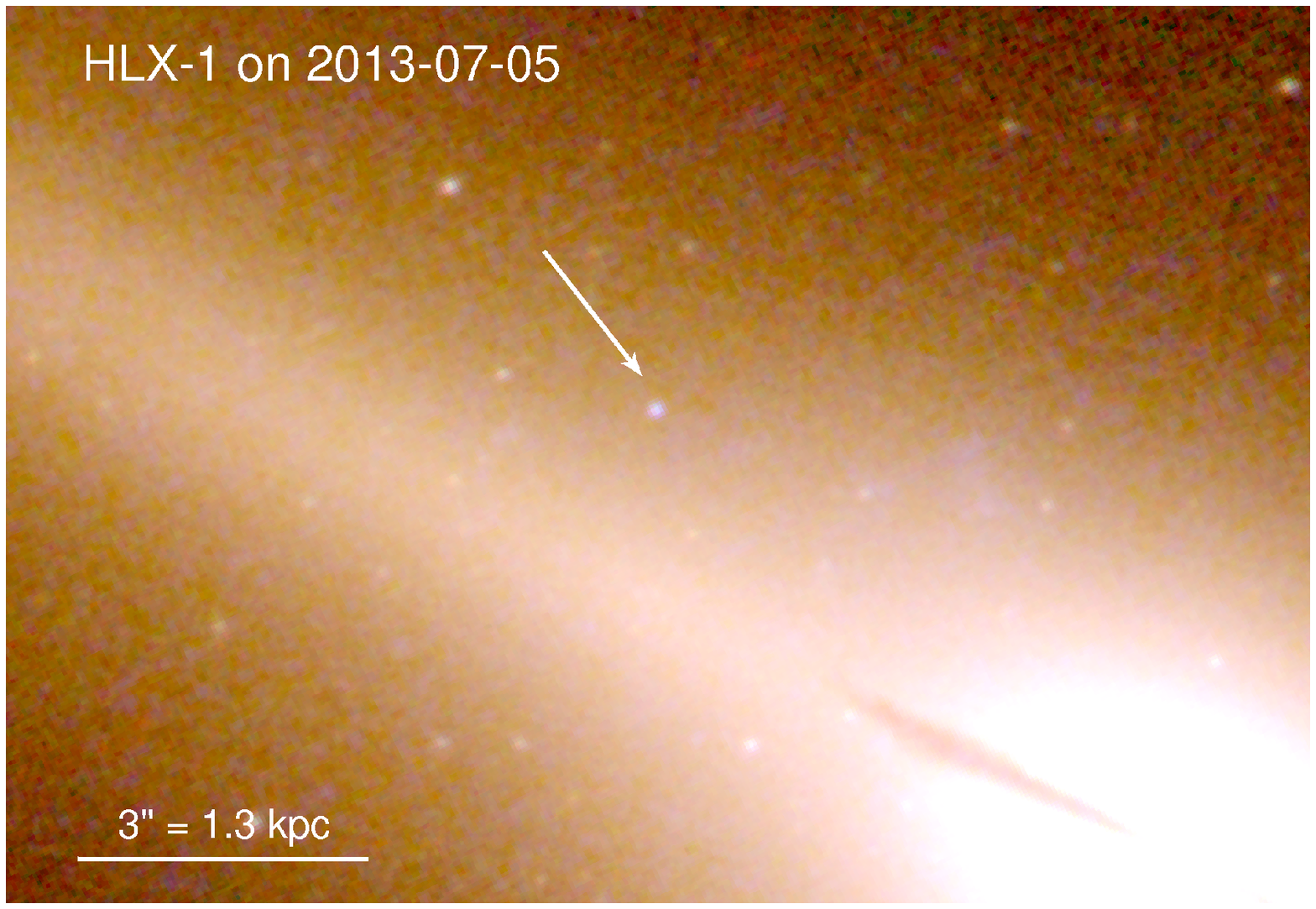,width=8.8cm,angle=0}\\
\vspace{-0.2cm}
\caption{HLX-1 appears as a point-like blue source (marked by an arrow) projected in front of the galaxy ESO\,243-49. Top panel: {\it HST}/WFC3 image from 2010 September 23 (absolute brightness $M_V \approx -11$ mag); red corresponds to the F775W filter, green to the F555W filter, and blue to the F300X filter. Bottom panel: same as in the top panel, for the 2013 July 5 image (absolute brightness $M_V \approx -10$ mag). In both images, north is up and east to the left.}
\label{fig1}
\end{figure}

\begin{figure}
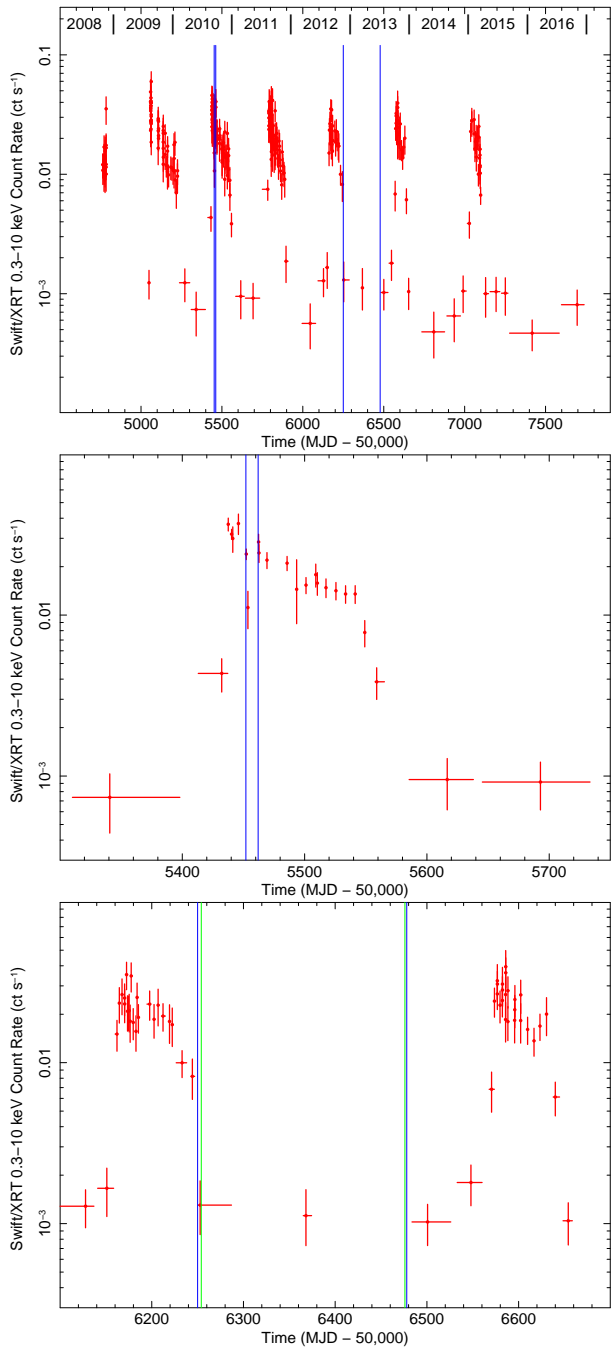

\psfig{figure=hlx1_lc_full.eps,width=5.9cm,angle=270}\\
\hspace{-0.6cm}
\psfig{figure=hlx1_xrt_lc_zoom1.eps,width=5.9cm,angle=270}\\
\hspace{-0.6cm}
\psfig{figure=hlx1_xrt_lc_zoom2.eps,width=5.9cm,angle=270}
\caption{Top panel: {\it Swift}/XRT snapshot lightcurve in the 0.3--10 keV band, binned by signal-to-noise $>$ 3 (updated to the end of 2016). Blue vertical lines mark the epochs of the {\it HST} observations. Middle panel: zoomed-in view of the 2010 outburst, binned by XRT observation and signal-to-noise $>$ 3. Bottom panel: same for the 2012 and 2013 outbursts; blue lines are the {\it HST} observations, green lines are the two {\it XMM-Newton} observations discussed in this work.}
\label{fig2}
\end{figure}

\begin{table*}
	\centering
	\caption{Summary of the {\it HST} observation log and results.}	
        \label{tab:hst_log_table}
	\begin{tabular}{lccccccr} 
		\hline
		Date & Days after peak & Instrument & Filter 
                & Central Wavelength & Exposure Time & Brightness & Brightness\\
		 &  & & & (\AA) & (s) & (Vegamag) & (ABmag)\\
		\hline
		2010-09-13 & 6  & ACS SBC   & F140LP & 1528.0  & 2480 & $22.13 \pm 0.15$ & $24.30 \pm 0.15$\\
		2010-09-23 & 16 & WFC3 UVIS & F300X  & 2814.8  & 1710 & $22.60 \pm 0.05$ & $24.07 \pm 0.05$\\
                           &    &           & F390W  & 3922.9  &  712 & $23.83 \pm 0.05$ & $24.05 \pm 0.05$\\ 
                           &    &           & F555W  & 5308.2  &  742 & $24.04 \pm 0.05$ & $24.02 \pm 0.05$\\ 
                           &    &           & F775W  & 7647.6  &  740 & $23.71 \pm 0.05$ & $24.10 \pm 0.05$\\ 
		           &    & WFC3 IR   & F160W  & 15369.2 &  806 & $23.6 \pm 0.3$ & $24.9 \pm 0.3$\\  
                \hline
		2012-11-19 & 72 & ACS SBC   & F140LP & 1528.0  & 2428 & $22.53 \pm 0.15$ & $24.70 \pm 0.15$\\
		           &    & WFC3 UVIS & F300X  & 2814.8  & 1001 & $23.2 \pm 0.1$ & $24.7 \pm 0.1$\\
                           &    &           & F336W  & 3354.8  &  983 & $23.5 \pm 0.1$ & $24.7 \pm 0.1$\\
                           &    &           & F390W  & 3922.9  & 1068 & $24.33 \pm 0.05$ & $24.55 \pm 0.05$\\  
                           &    &           & F555W  & 5308.2  & 1045 & $24.57 \pm 0.05$ & $24.55 \pm 0.05$\\  
                           &    &           & F621M  & 6218.9  & 1065 & $24.40 \pm 0.05$ & $24.55 \pm 0.05$\\ 
                           &    &           & F775W  & 7647.6  & 1040 & $24.11 \pm 0.05$& $24.50 \pm 0.05$\\ 
		           &    & WFC3 IR   & F105W  & 10551.0 & 1209 & $24.1 \pm 0.2$ & $24.7 \pm 0.2$\\ 
		           &    &           & F160W  & 15369.2 & 1209 & $23.7 \pm 0.3$ & $25.0 \pm 0.3$\\
                \hline
		2013-07-05 & 300 & ACS SBC   & F140LP & 1528.0  & 2428 & $23.83 \pm 0.15$ & $26.00 \pm 0.15$\\
		           &     & WFC3 UVIS & F300X  & 2814.8  & 1004 & $24.1 \pm 0.1$ & $25.6 \pm 0.1$\\
                           &     &           & F336W  & 3354.8  &  980 & $24.2 \pm 0.1$ & $25.4 \pm 0.1$\\
                           &     &           & F390W  & 3922.9  & 1074 & $25.2 \pm 0.2$ & $25.4 \pm 0.1$\\ 
                           &     &           & F555W  & 5308.2  & 1039 & $25.0 \pm 0.1$ & $25.0 \pm 0.1$\\ 
                           &     &           & F621M  & 6218.9  & 1077 & $24.85 \pm 0.05$ & $25.00 \pm 0.15$\\  
                           &     &           & F775W  & 7647.6  & 1028 & $24.51 \pm 0.15$ & $24.90 \pm 0.15$\\  
		           &     & WFC3 IR   & F105W  & 10551.0 & 1209 & $24.3 \pm 0.2$ & $24.9 \pm 0.2$\\ 
		           &     &           & F160W  & 15369.2 & 1209 & $23.8 \pm 0.3$ & $25.1 \pm 0.3$\\
		\hline
	\end{tabular}
\end{table*}

\section{Optical/UV and X-ray study}

\subsection{Setting the problem: distinguishing accretion disk and stellar contributions}
The point-like optical source discovered at the position of HLX-1 \citep{soria10}, in a region devoid of any other bright X-ray or optical point-like sources, might in principle be a chance coincidence, rather than being physically associated with the hyperluminous X-ray source. If it were a chance association, the detection of a redshifted H$\alpha$ emission line from the optical counterpart \citep{wiersema10,soria13b} suggests that the optical source is at $\approx$100 Mpc while the X-ray source might be a lower-luminosity foreground X-ray binary. This is statistically implausible, and would also make the disk temperature and luminosity of HLX-1 no longer consistent with canonical BH accretion states \citep{farrell09,yan15}. Further strong evidence of a physical connection between X-ray and optical emission comes from their variability properties, as discussed below. 

The physical interpretation of the optical/UV emission has been a topic of intense debate. The source is too bright ($M_V \approx -11$ mag in outburst: \citealt{farrell12}) to be an individual O star or even a group of few O stars. Two alternative possibilities were proposed \citep{farrell12,soria12}, based on combined fits of the X-ray and optical/UV spectrum in outburst: either the optical emission is dominated by a young, massive ($M_{\ast} \sim 10^6 M_{\odot}$) star cluster, or it comes mostly from the X-ray-irradiated accretion disk, supplemented by an older stellar population (age $> 10$ Gyr) to account for an observed near-infrared excess.

Different formation and evolution scenarios for the candidate IMBH are supported by either interpretation of the optical counterpart. If {\bf most of} the near-UV and blue emission is from a young star cluster, a recent, localized episode of intense star formation is required. There is no other evidence of recent star formation in the halo of ESO\,243-49, but the young cluster could be the nucleus of a recently accreted gas-rich dwarf galaxy \citep{farrell12,mapelli12,mapelli13a,mapelli13b}. A young, massive star cluster could be a suitable location for the formation and growth of an IMBH \citep{pz02,gurkan04}. 
Instead, if the near-UV and blue light is mostly reprocessed thermal emission from an irradiated accretion disk, there is much less need for a massive supply of gas and substantial recent star formation in the star cluster. If the donor star is a red giant or asymptotic giant branch star, or a blue straggler, the cluster may consist entirely of an old population; alternatively, the cluster may contain a few young, massive stars near the centre (including the IMBH donor), perhaps formed from a small amount of gas swept up from the interstellar medium \citep{li16,conroy11}. Moreover, if the near-UV and blue emission turn out to be from the irradiated disk, we can use that observed flux to obtain a characteristic size of the emitting region, and from that, constrain the size of the BH Roche lobe and the binary separation.

In the young cluster scenario, the optical/UV emission is not expected to show substantial variations on timescales shorter than the evolution or the dynamical timescale of the cluster. In the irradiated disk scenario, brightness variations in X-rays and optical are naturally expected, as a system could get into an outburst phase or retreat into quiescence.
Very Large Telescope (VLT) observations taken just before and during the rise of the 2012 outburst showed an increase in the visual brightness of $\Delta V = 1.8 \pm 0.4$ mag between near-quiescence (365 days after the peak of the previous outburst) and outburst peak \citep{webb14}. In the $R$ band, a comparison of observations taken with the {\it Hubble Space Telescope} ({\it HST}), VLT, Gemini and Magellan at different times over 2009--2012 suggests a brightness change of $\Delta R = 0.9 \pm 0.4$ mag between outburst and quiescence \citep{farrell14,webb14}; however, this result may be affected by systematic errors in the conversion between different filter bands. Moreover, the diffuse emission from the old (red) stellar population in the halo of ESO\,243-49 substantially reduces the detection significance of the HLX-1 counterpart from ground-based telescopes, especially towards quiescence.

\subsection{{\it HST}, {\it XMM-Newton} and {\it Swift} data analysis}

In this work, we present and compare the photometric results from three sets of {\it HST} observations, taken with the Wide Field Camera 3 (UVIS and IR apertures), on 2010 September 23 ($\approx$16 days after outburst peak), on 2012 November 19 ($\approx$72 days after peak) and on 2013 July 5--6 ($\approx$300 days after peak) (Table 1). Ultraviolet observations were also taken on 2010 September 13, 2012 November 19 and 2013 July 5 with the Advanced Camera for Surveys Solar Blind Channel (ACS SBC). To model the optical/UV data, we fit them simultaneously with representative X-ray spectra taken as close as possible to the optical observations, and/or at a similar level of X-ray luminosity. For the 2012 {\it HST} data, we used a 54-ks {\it XMM-Newton} observation taken on November 23. For the 2013 {\it HST} data, we used a 141-ks {\it XMM-Newton} observation taken on July 3. For the 2010 {\it HST} data, no contemporaneous {\it XMM-Newton} observations are available, and individual {\it Swift} observations are too short to provide a meaningful constraint to the X-ray parameters. Therefore, we determined the {\it Swift} X-Ray Telescope (XRT) count rate at the time of the 2010 {\it HST} observations ({\it Swift} observation of 2010 September 23), and then stacked all {\it Swift} observations across the 6 observed outbursts, that have an observed count rate within a factor of 2 of the count rate at the {\it HST} epoch\footnote{We can build this average spectrum because we also verified that all {\it Swift} observations in that range of count rates ($\approx$0.012--0.05 ct s$^{-1}$) have similar colours in a hardness-intensity diagram, consistent with the disk-dominated high/soft state; see also \cite{yan15}.}.


\begin{figure*}
\psfig{figure=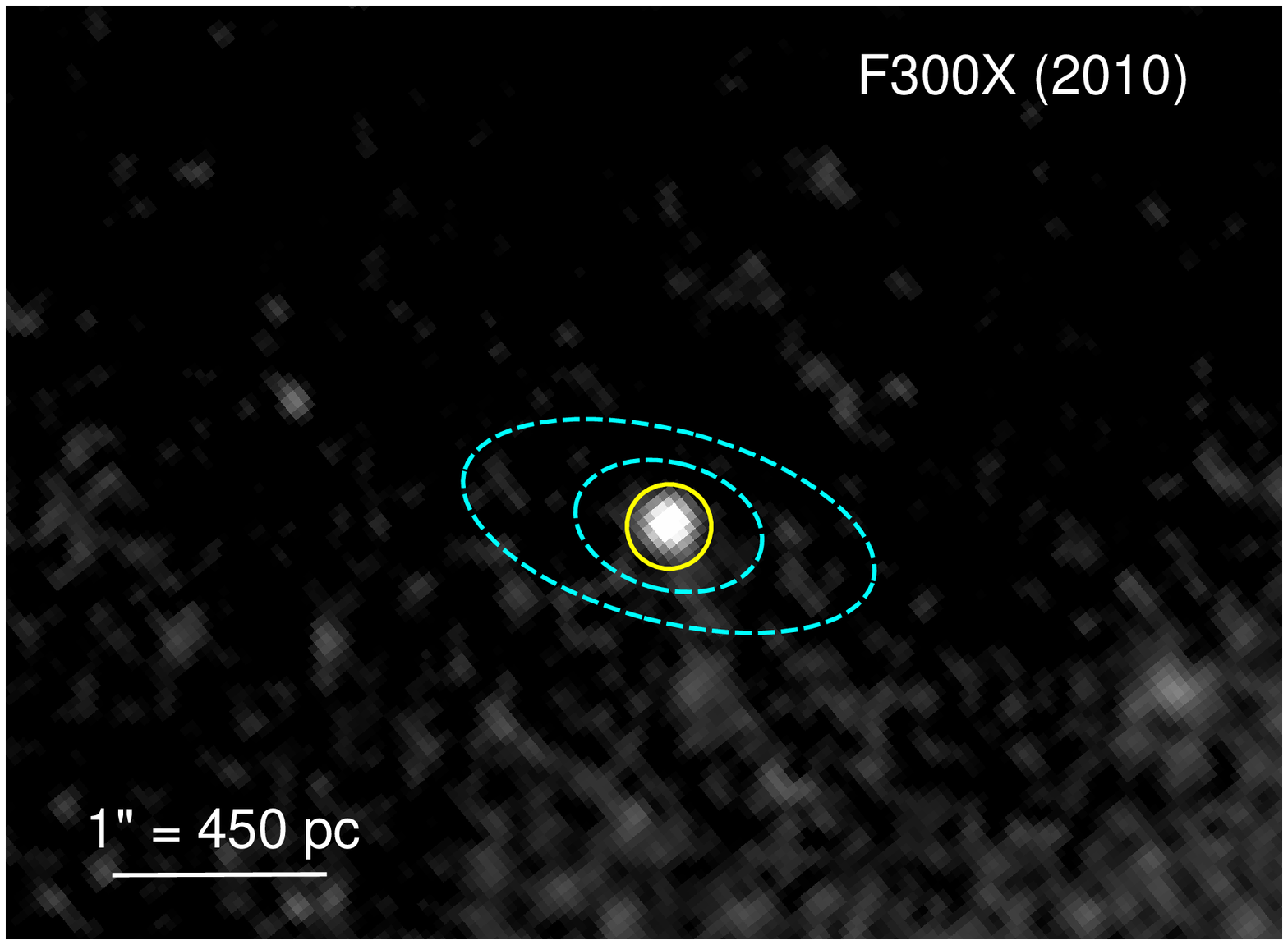,width=6.0cm,angle=0}
\hspace{-0.4cm}
\psfig{figure=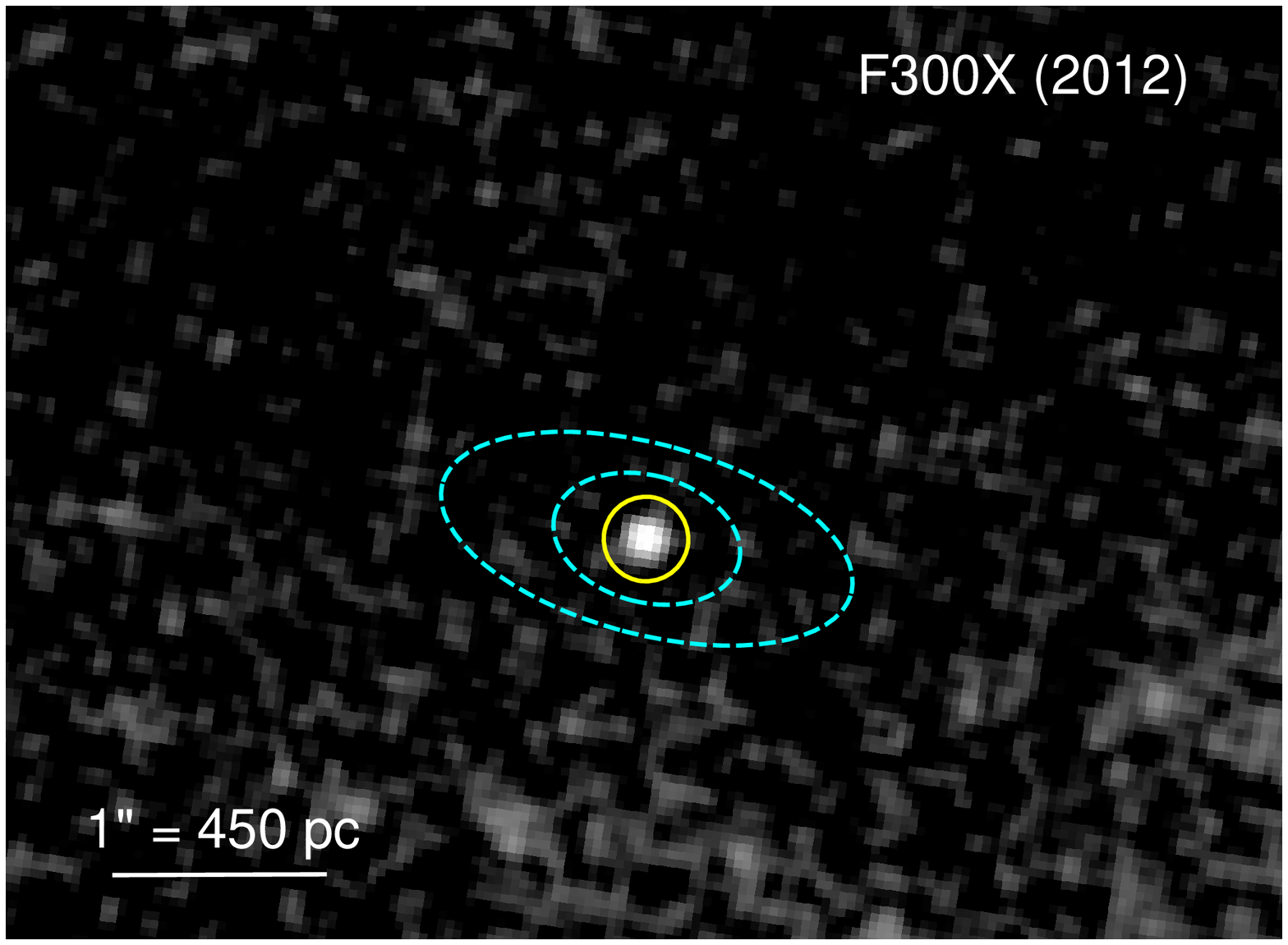,width=6.0cm,angle=0}
\hspace{-0.4cm}
\psfig{figure=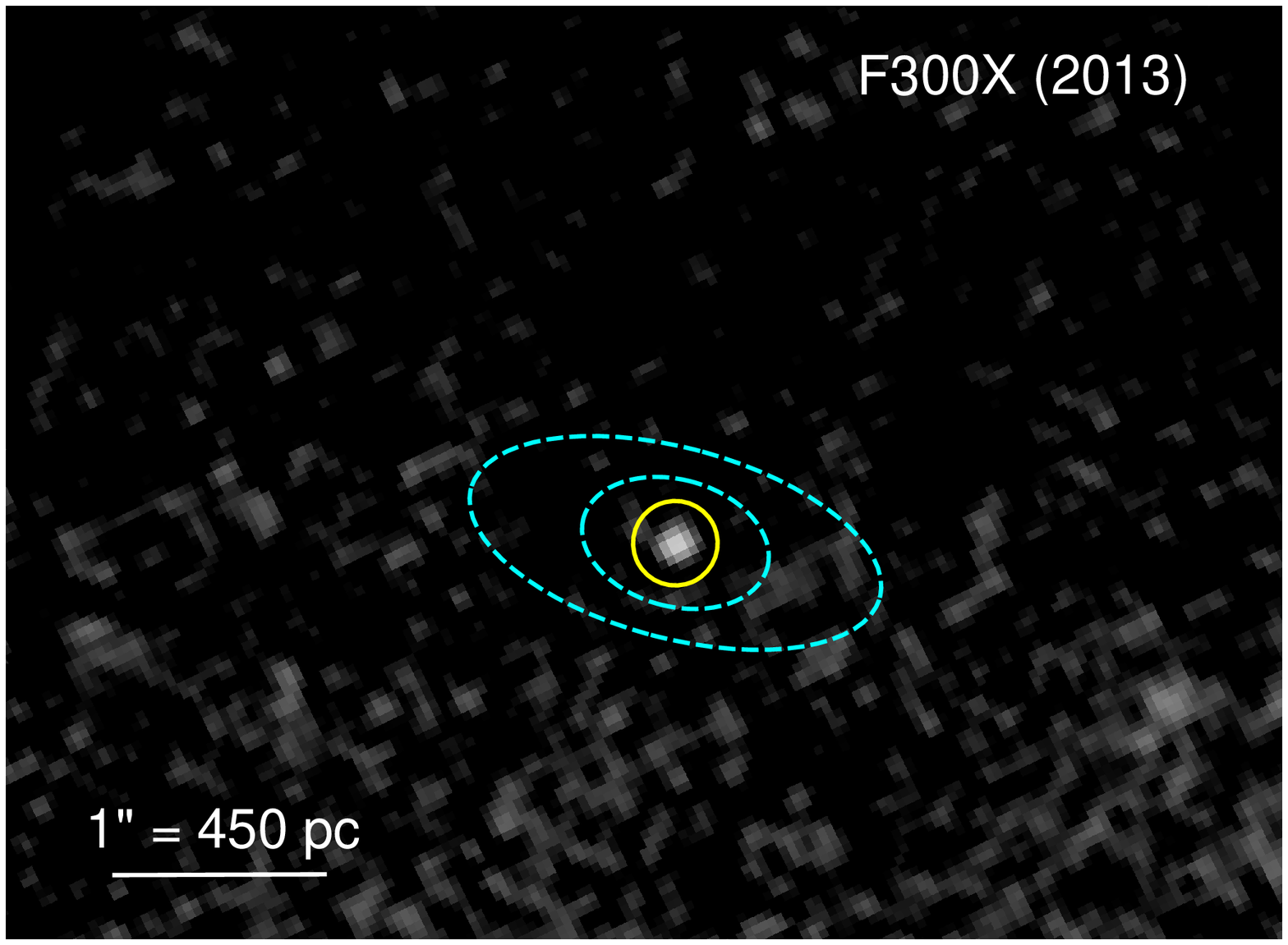,width=6.0cm,angle=0}\\

\vspace{-0.1cm}
\psfig{figure=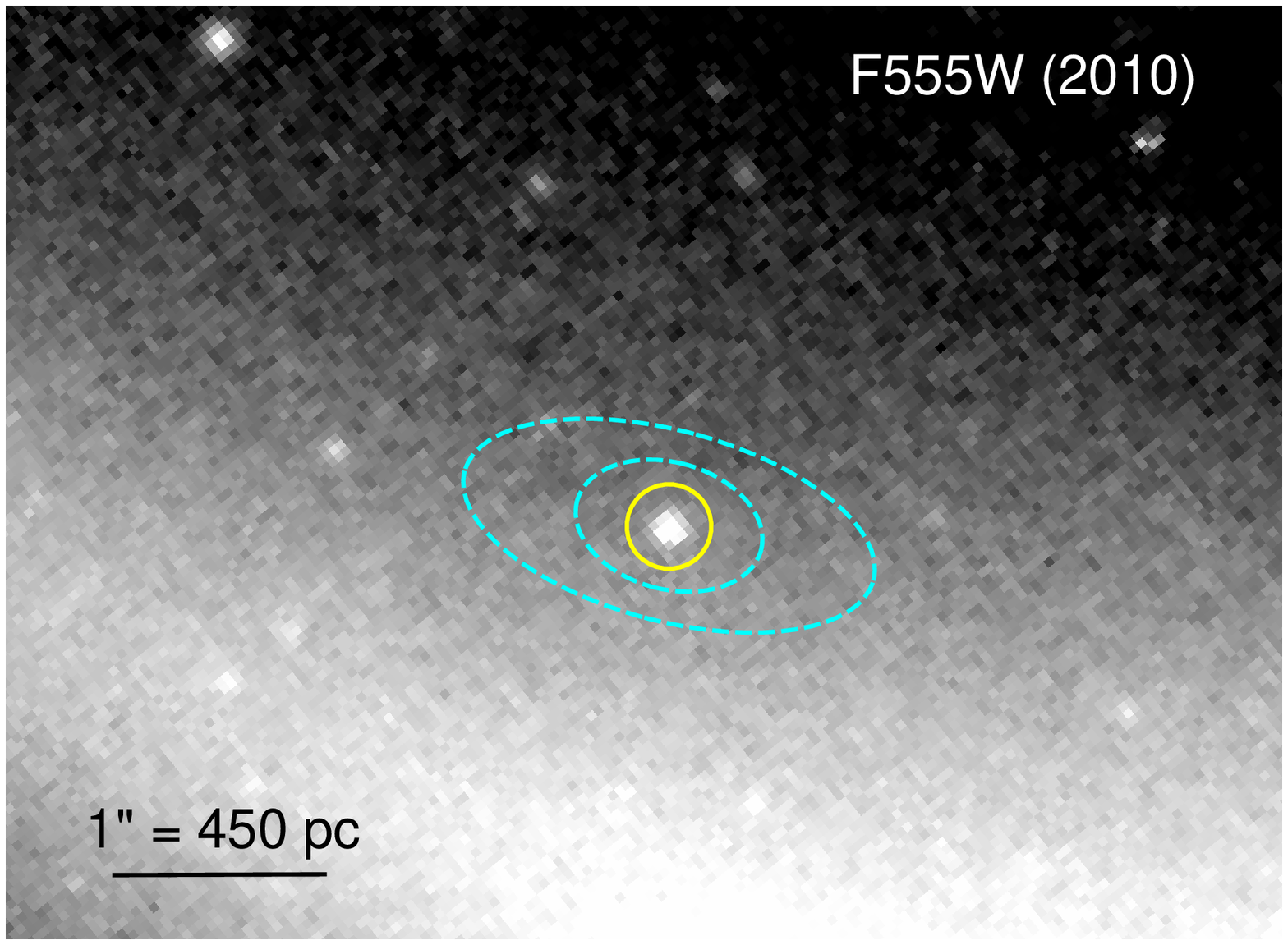,width=6.0cm,angle=0}
\hspace{-0.4cm}
\psfig{figure=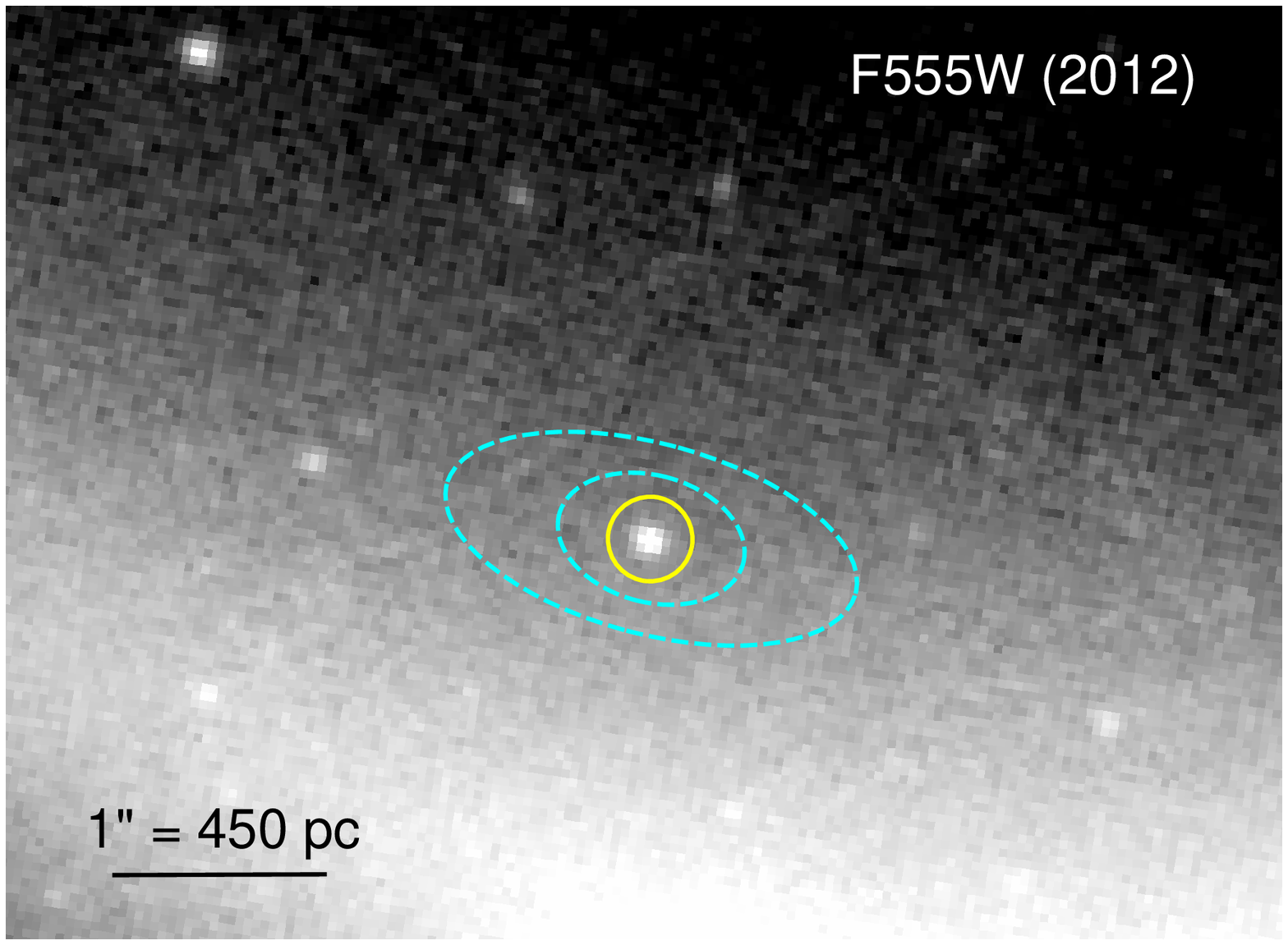,width=6.0cm,angle=0}
\hspace{-0.4cm}
\psfig{figure=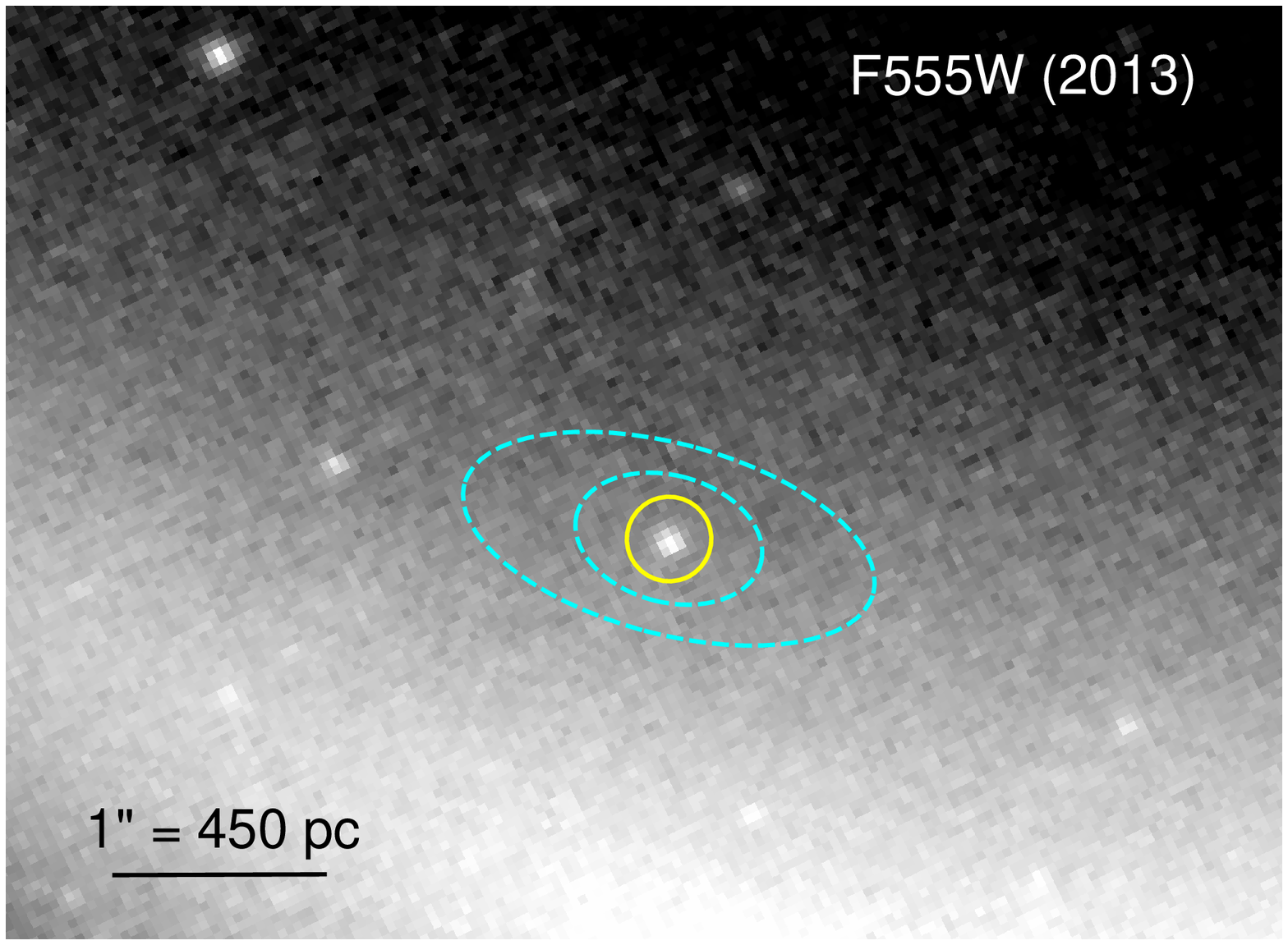,width=6.0cm,angle=0}\\

\vspace{-0.1cm}
\psfig{figure=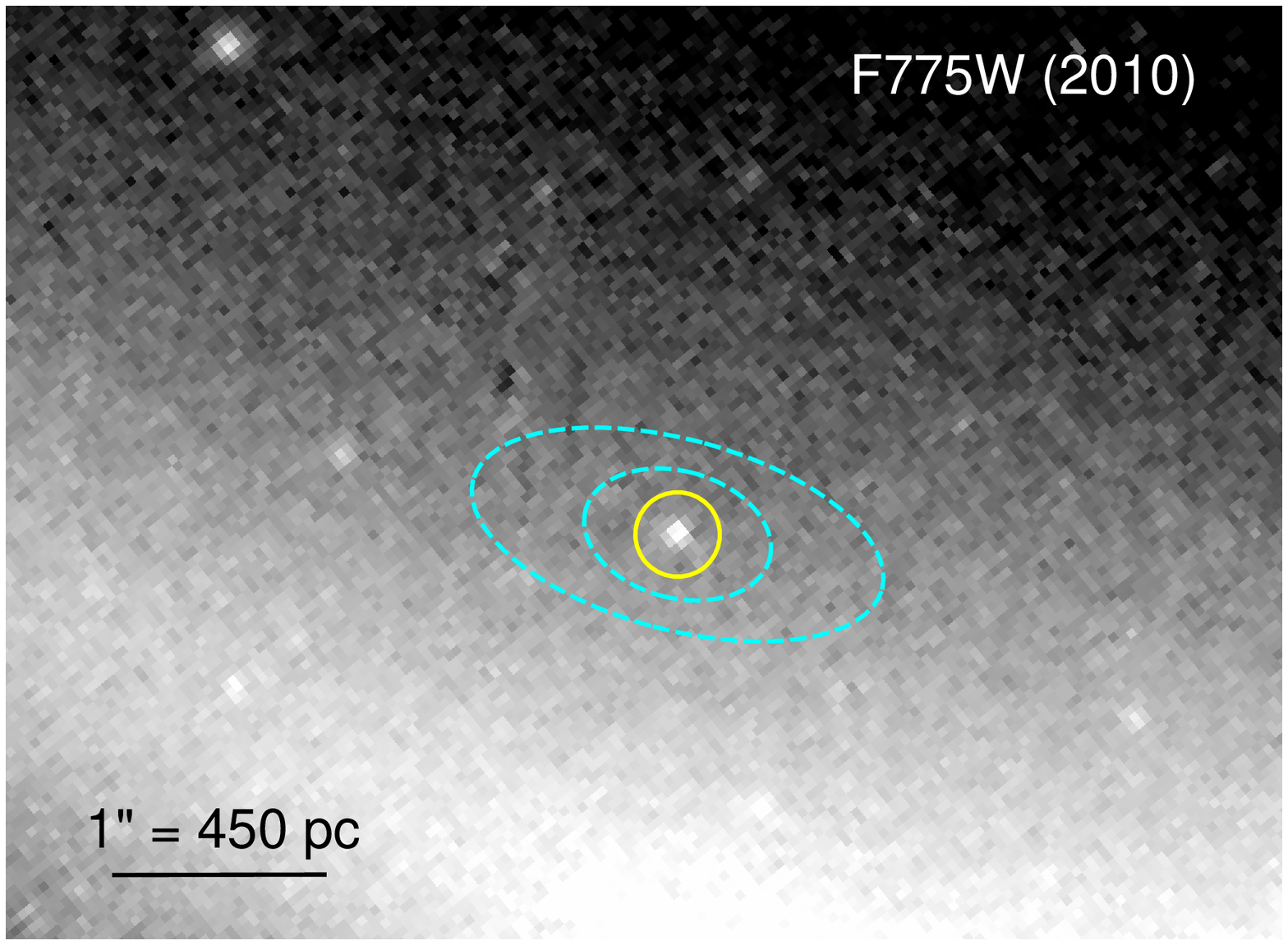,width=6.0cm,angle=0}
\hspace{-0.4cm}
\psfig{figure=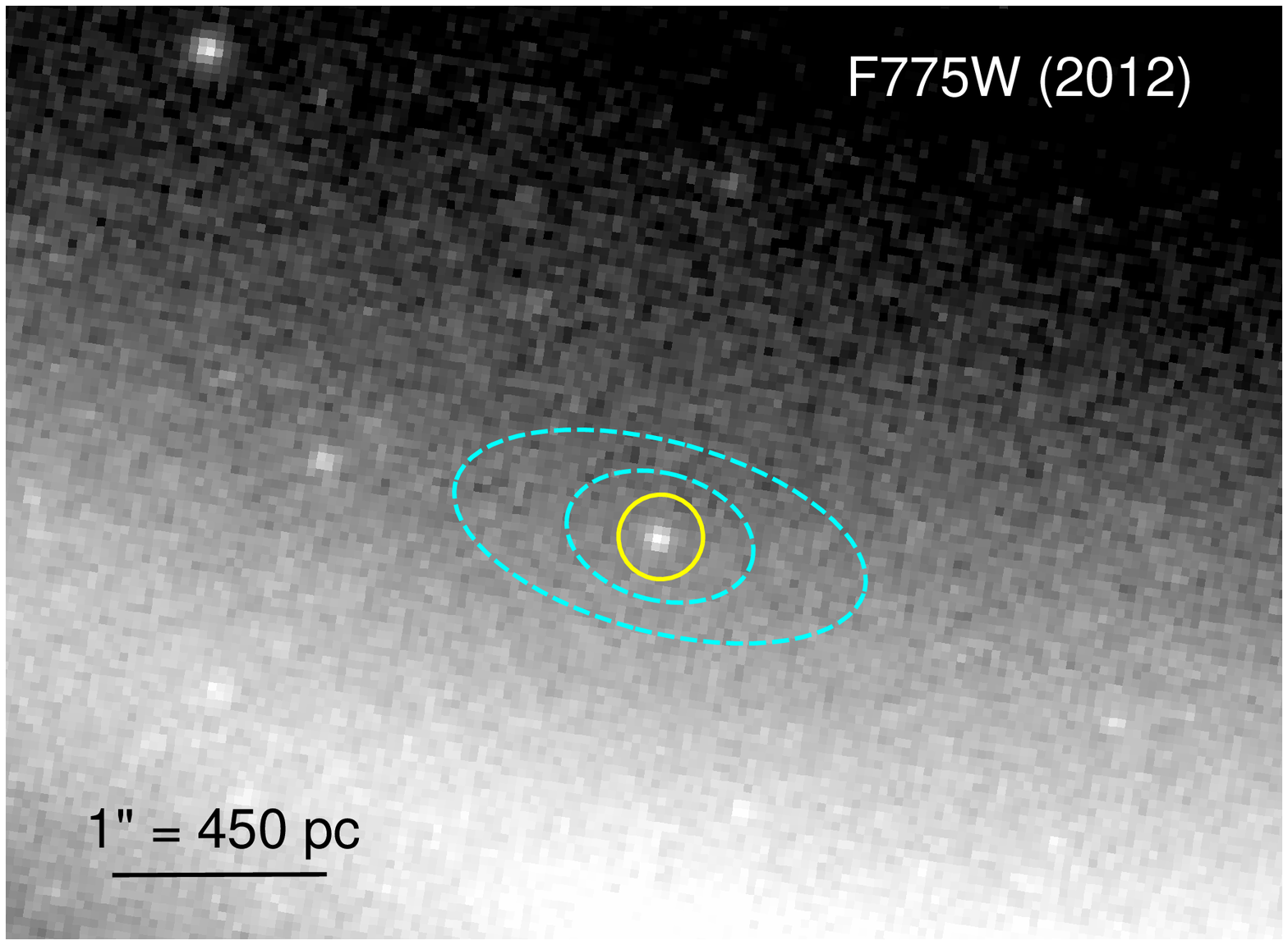,width=6.0cm,angle=0}
\hspace{-0.4cm}
\psfig{figure=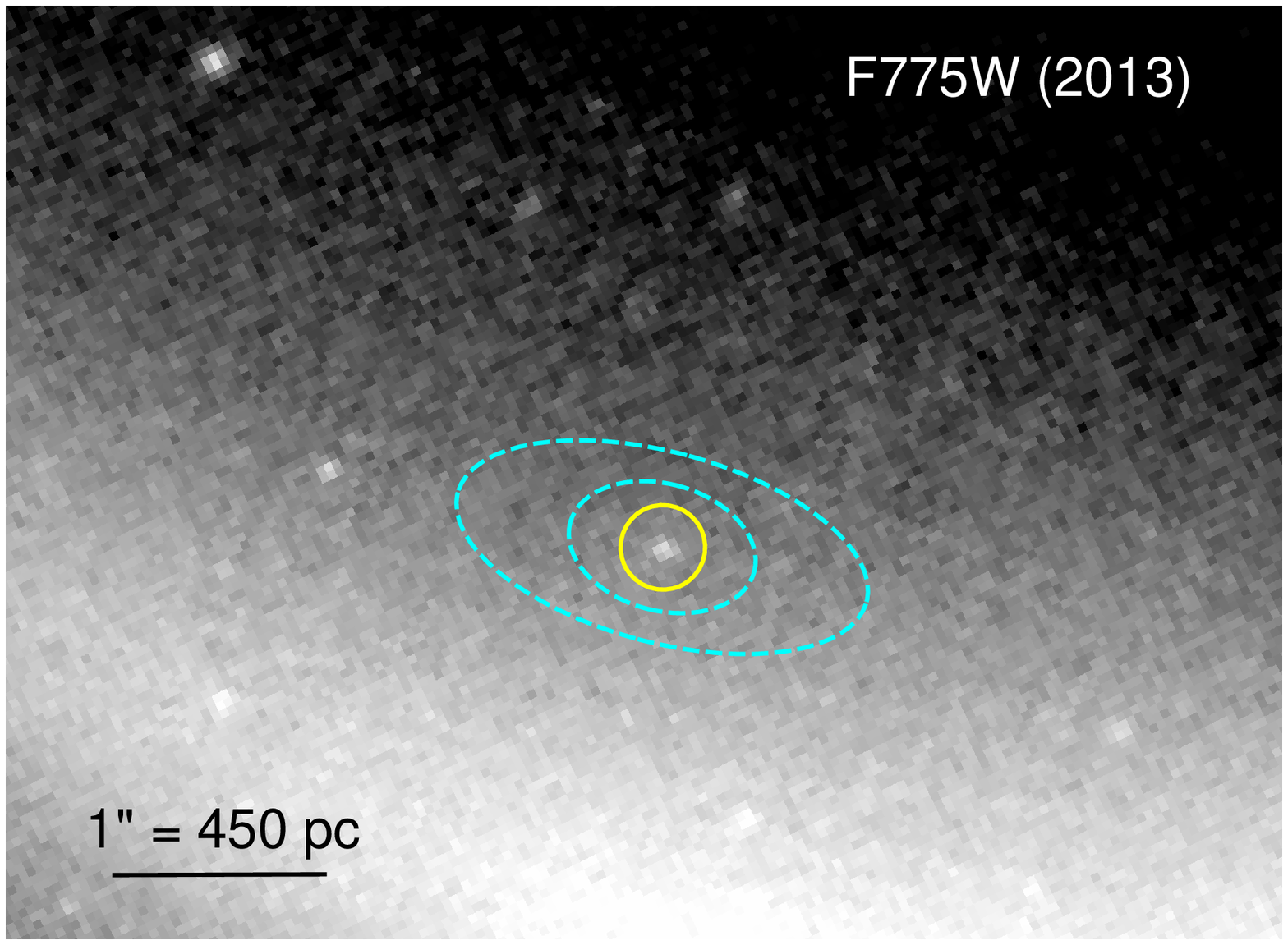,width=6.0cm,angle=0}
\caption{Zoomed-in view of HLX-1 from the {\it HST}/WFC3 images, at different epochs and for a selection of three different filters. North is up and east to the left. In each panel, the yellow circle (0".2 radius) is the source extraction region in our photometric analysis, while the background was extracted from the region between the two dashed ellipses. Top row: F300X filter; middle row: F555W filter; bottom row: F775W filter. In each row, we used the same greyscale for the three epochs. The luminosity decline between the 2010 and 2013 datasets was 1.5 mag in the F300X filter, 1.0 mag in the F555W filter, and 0.8 mag in the F775W filter.}
\label{fig3}
\end{figure*}


For the {\it HST} data, we downloaded the calibrated images (drizzled files) from NASA's Mikulski Archive for Space Telescopes (MAST). 
HLX-1 is a well-isolated source, with no risk of confusion with other nearby point-like sources. Therefore we used aperture photometry to measure its brightness, with standard packages such as {\small {SAOImage DS9}} Version 7.4 and {\small{IRAF}} Version 2.16. Particularly in the redder filters, proper background subtraction is crucial because of the strong unresolved emission from the old stellar population in ESO\,243-49 and the gradient of such emission. For the source, we used a circular extraction region of radius 0''.2 (Figure 3). For the background, we used an elliptical annulus oriented at a parallactic angle (north through east) of 75$^{\circ}$, that is approximately parallel to the isophotes of ESO\,243-49 around the projected location of HLX-1. This was done to reduce the gradient of the unresolved emission in the background region. The semi-major axes of the outer background annulus were $1''$ and $0''.45$; those of the inner (exclusion) annulus were $0''.45$ and $0''.3$. For each filter and each epoch, we measured the background-subtracted count rate of HLX-1 within the $0''.2$ extraction region, and converted it to a $0''.4$ count rate. To do so, we selected isolated, brighter point-like sources in the same chip, and determined the ratio of the count rates from a $0''.2$ and $0''.4$ radius; as a safety check, we also compared our empirical values with the aperture correction values tabulated in the WFC3 and ACS online handbooks, and found them consistent. We then converted the $0''.4$ count rates to Vega and AB magnitudes, using the Zeropoint tables provided online by the Space Telescope Science Institute. Finally, we used the {\small FTOOLS} \citep{blackburn95} task {\it flx2xsp} to convert the observed {\it HST} flux densities into standard PHA files with their associated response files, which can be displayed and fitted in {\small XSPEC} \citep{arnaud96}.




For {\it XMM-Newton}, we downloaded the data from NASA's High Energy Astrophysics Science Archive Research Center (HEASARC) archive: we used ObsID 0693060401 (PI: S.~Farrell) from 2012 November 23, and ObsID 0693060301 (PI: S.~Farrell) from 2013 July 4--5. We reprocessed the European Photon Imaging Camera (EPIC) MOS and pn observation data files with the Science Analysis System (SAS) version 14.0.0. 
We checked for exposure intervals with high particle background, and removed them from the analysis; the good-time-interval was 49 ks for the 2012 dataset and 112 ks for the 2013 dataset. We extracted the source photons from a circular region with a radius of $30''$; the background photons were obtained from nearby regions located at similar distances from readout nodes, and avoiding chip gaps. We used the {\small SAS} task {\it xmmselect} to select single and double events (pattern 0--4 for pn and 0--12 for MOS1 and MOS2), and filtered them with the standard criteria FLAG $= 0$ \& \#XMMEA\_EP for the pn and \#XMMEA\_EM for the MOS. We built response and ancillary response files with the {\small SAS} tasks {\it rmfgen} and {\it arfgen}. Finally, to increase the signal-to-noise ratio, we combined the pn and MOS spectra with {\it epicspeccombine}, creating an average EPIC spectrum for each of the two epochs. We grouped the two combined spectra to a minimum of 20 counts per bin;  coincidentally, this gives us $\approx$3 bins per spectral resolution element, keeping in mind that the spectral resolution of a combined EPIC spectrum is $\approx$100 eV at $\approx$0.3--2 keV. 
Finally, we fitted the 2012 and 2013 EPIC spectra with {\small XSPEC} version 12.8.2 \citep{arnaud96}, together with the corresponding {\it HST} spectra.

For the {\it Swift}/XRT data, we used the online product generator \citep{evans09} to extract a lightcurve, determine the count rate on 2010 September 23, and build a stacked spectrum with all the XRT observations between 2009 and 2015 with a count rate within a factor of 2 of the reference one. The stacked spectrum comprises 101 snapshot observations for a total exposure time of 66 ks. A similar stacking technique was previously used by other authors ({\it e.g.}, \citealt{yan15,soria11}) to model the characteristic high/soft, intermediate and low/hard state spectra. We verified that a stacked spectrum including only data from the 2010 outburst (similar to what was done by \citealt{farrell12} for their broad-band modelling of X-ray and optical/UV data) produces similar results, but at lower signal-to-noise, which makes it harder to constrain the disk parameters. 

\begin{figure}
\hspace{-0.4cm}
\psfig{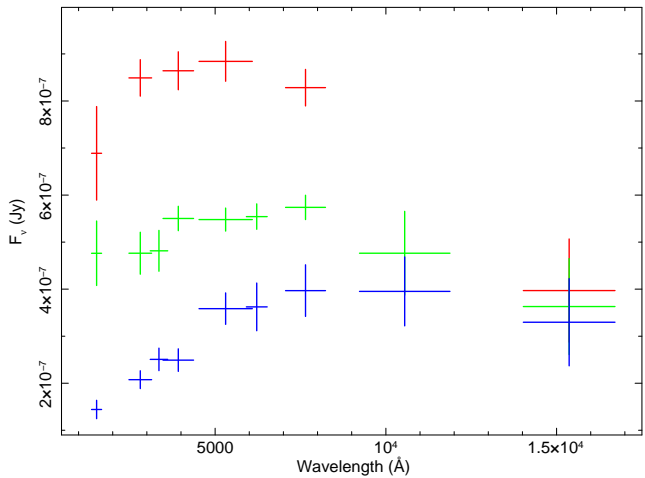}
\caption{Red datapoints: optical flux densities in the 2010 {\it HST} dataset (corresponding to the high/soft state); green datapoints: optical flux densities in the 2012 dataset (intermediate state); blue datapoints: optical flux densities in the 2013 dataset (low/hard state).}
\label{fig4}
\end{figure}

\subsection{Main results}

The immediate result of our comparison between the three {\it HST} observations is the sharp decline in the bluest part of the optical/UV spectrum going from the high/soft to the low/hard X-ray state; instead, there is only a small flux decrease in the redder bands (Table 1 and Figure 4). It was already known \citep{farrell12,soria12,farrell14} that a single irradiated disk is not sufficient to reproduce the 2010 {\it HST} optical spectrum: either an additional near-IR component (possibly an old stellar population), or an additional near-UV component (possibly a young stellar population), or both, are needed. For each of the three {\it HST} epochs, we applied a code developed by \cite{mucciarelli07} and \cite{patruno08}, upgraded to calculate the magnitudes of an irradiated disk in the {\it HST} photometric bands (see also \citealt{mapelli13a}): like for the 2010 data, we verified that at least one additional component is required also for the 2012 and 2013 data. By comparing the three datasets, we have now proved the correlation between near-UV emission and the level of X-ray irradiation. Therefore, the near-UV emission in outburst cannot be dominated by a young stellar population, because of its variability; the near-UV luminosity in the faintest observed epoch (2013) sets the upper limit to the constant contribution of young stars. On the other hand, the near-IR component can be dominated by constant emission from an old stellar population.

\begin{table*}
    \centering
    \begin{threeparttable}
    \caption{Spectral parameters of our X-ray/optical spectral fitting for the three epochs of {\it HST} observations. The model (``Model 1") is {\it redden}$_{\rm {los}}$ $\times$ {\it redden} $\times$ {\it TBabs}$_{\rm {los}}$ $\times$ {\it TBabs} $\times$ ({\it diskir} $+$ {\it bbodyrad}) $+$ {\it TBabs}$_{\rm {los}}$ $\times$ {\it mekal} (the last term accounts for the contamination of thermal emission from ESO\,243-49). Errors are 90\% confidence limits for 1 interesting parameter.}
\begin{tabular}{llrrr}
        \hline
        Component & Parameter & \multicolumn{3}{c}{Epoch}\\
        &&\multicolumn{1}{r}{2010 Sep} & \multicolumn{1}{r}{2012 Nov} & \multicolumn{1}{r}{2013 Jul}\\
        \hline
        {\it redden}$_{\rm los}$  & $E(B-V)_{\rm los}$ (mag) & [0.013] & [0.013] & [0.013]\\[2pt]
        \hline
        {\it redden}   & $E(B-V)$ (mag) & $< 0.12$ & $<0.12$ & $< 0.11$ \\[2pt]
        \hline
        {\it TBabs}$_{\rm los}$ & $N_{\rm H,los}$ ($10^{20}$ cm$^{-2}$) & [2.0] & [2.0] &  [2.0] \\[2pt]
        \hline
        {\it TBabs} & $N_{\rm H}$  ($10^{20}$ cm$^{-2}$)   & $<1.3$ & $< 1.4$ & $<2.0$\\[2pt]
        \hline
        {\it diskir}&$kT_{\rm in}$ (keV)& \U{0.22}{0.01}{0.01} & \U{0.083}{0.037}{0.040}& $< 0.073$\\[2pt]
        & $\Gamma$ & \U{1.8}{2.1}{0.6}&\U{2.43}{0.15}{0.15} & \U{2.04}{0.26}{0.20}\\[2pt]
        &$kT_{\rm e}$ (keV) & [100]  & [100] & [100]\\[2pt]
        & $L_{\rm c}/L_{\rm d}$ & \U{0.12}{1.35}{0.09}& \U{2.1}{\ast}{0.9} & \U{16.8}{\ast}{15.4} \\[2pt]
        & $f_{\rm in}$ & [0.1] & [0.1] & [0.1] \\[2pt]
        & $r_{\rm irr}$ & [1.2] & [1.2] & [1.2] \\[2pt]
        & $f_{\rm out}$ $(10^{-3})$ & \U{6.6}{1.5}{3.5} &  \U{43}{\ast}{16} & \U{45}{23}{15} \\[2pt]
        & log($R_{\rm out}$) & \U{3.68}{0.05}{0.06} & \U{3.27}{0.30}{0.24} & $< 3.16$ \\[2pt]
        &$K$\tnote{{\it{a}}}
        & \U{17.6}{12.2}{4.4}& \U{39.5}{139}{28.5}& $> 3.0$\\[2pt]
        \hline
        {\it bbodyrad} & $T_{\rm bb}$ (K) & [5270] & [5270] & \U{5270}{300}{400}\\[2pt]
        & $N_{\rm bb}$ ($10^9$)\tnote{{\it{b}}} & [3.9] & [3.9] & \U{3.9}{0.7}{0.9} \\[2pt]
        \hline
        {\it mekal} & $kT_{\rm mk}$ (keV) & [0.45] & [0.45] & \U{0.45}{0.14}{0.13}\\[2pt]
          & $N_{\rm mk}$ ($10^{-6}$) & [1.1] & [1.1] & \U{1.1}{0.7}{0.5}\\[2pt]
        \hline
        & $f_{\rm X}$ ($10^{-14}$  erg cm$^{-2}$ s$^{-1}$)\tnote{{\it{c}}}  & \U{57}{3}{3}& \U{7.2}{0.4}{0.2} & \U{2.6}{0.2}{0.2}\\[2pt]
        & $L_{\rm X} \cos \theta$  ($10^{40}$ erg s$^{-1}$)\tnote{{\it{d}}} & \U{33.1}{1.8}{1.9} &\U{4.1}{0.3}{0.2}& \U{2.8}{0.4}{0.2} \\[2pt]        
        \hline
        & $f_{\rm O}$ ($10^{-14}$  erg cm$^{-2}$ s$^{-1}$)\tnote{{\it{e}}}  & \U{2.4}{0.1}{0.1}& \U{1.4}{0.1}{0.1} & \U{0.60}{0.02}{0.02} \\[2pt]
        & $f_{\rm R}$  ($10^{-14}$ erg cm$^{-2}$ s$^{-1}$)\tnote{{\it{f}}} & [0.18] & [0.18] & \U{0.18}{0.04}{0.05} \\[2pt]
        \hline
        & $R_{\rm in} \, \sqrt{\cos \theta}$ ($10^8$ cm) & \U{46}{14}{6} & \U{68}{61}{32} & $> 20$ \\[2pt]
        \hline
        & $\chi^2_{\nu}$ & 1.29 (91.3/71) & 0.88 (64.3/73) & 0.75 (61.5/82) \\[2pt]
        \hline
    \end{tabular}
        \begin{tablenotes}
      {\footnotesize{
      \item[{\it {a}}]{$K \equiv \left(r_{\rm in}/{\rm{km}}\right)^2\,\left(10\,{\rm {kpc}}/d \right)^2 \, \cos \theta$.}
      \item[{\it {b}}]{$N_{\rm bb} \equiv \left(R_{\rm bb}/{\rm{km}}\right)^2\,\left(10\,{\rm {kpc}}/d \right)^2$.}
      \item[{\it {c}}] Observed flux in the 0.3--10 keV band (not including the {\it mekal} component).
      \item[{\it {d}}]{Unabsorbed luminosity in the 0.3--10 keV band (not including the {\it mekal} component); $L_{\rm X} \equiv (2\pi \, d^2/\cos \theta)\, f_{\rm X}$ for the 2010 (high) and 2012 (intermediate) state, and $L_{\rm X} \equiv 4\pi \, d^2\, f_{\rm X}$ for the 2013 (low) state.
      \item[{\it {e}}] Total observed flux at $\lambda > 912$ \AA.
      \item[{\it {f}}] Observed flux that is attributed to a constant red component.
      }
      }}
    \end{tablenotes}
  \end{threeparttable}
    \vspace{0.3cm}
    \label{tab_diskir}
\end{table*}

\begin{table*}
    \centering
    \begin{threeparttable}
    \caption{Spectral parameters of our X-ray/optical spectral fitting for the three epochs of {\it HST} observations (``Model 2''). For 2010 and 2012, the model is {\it redden}$_{\rm los}$ $\times$ {\it redden} $\times$ {\it TBabs}$_{\rm los}$ $\times$ {\it TBabs} $\times$ ({\it diskir} $+$ {\it bbodyrad} $+$ {\it bbodyrad}) $+$ {\it TBabs}$_{\rm los}$ $\times$ {\it mekal} (the last term accounts for the contamination of thermal emission from ESO\,243-49). For 2013, the model is {\it redden}$_{\rm los}$ $\times$ {\it redden} $\times$ {\it TBabs}$_{\rm los}$ $\times$ {\it TBabs} $\times$ [({\it bbodyrad} $+$ {\it comptt}) $+$ {\it bbodyrad} $+$ {\it bbodyrad}) $+$ {\it TBabs}$_{\rm los}$ $\times$ {\it mekal}. Errors are 90\% confidence limits for 1 interesting parameter.}
\begin{tabular}{llrrr}
        \hline
        Component & Parameter & \multicolumn{3}{c}{Epoch}\\
        &&\multicolumn{1}{r}{2010 Sep} & \multicolumn{1}{r}{2012 Nov} & \multicolumn{1}{r}{2013 Jul}\\
        \hline
        {\it redden}$_{\rm los}$  & $E(B-V)_{\rm los}$ (mag) & [0.013] & [0.013] & [0.013] \\[2pt]
        \hline
        {\it redden}   & $E(B-V)$ (mag) & $< 0.10$ & $<0.15$ & [0.0] \\[2pt]
        \hline
        {\it TBabs}$_{\rm los}$ & $N_{\rm H,los}$ ($10^{20}$ cm$^{-2}$) & [2.0] & [2.0] &  [2.0] \\[2pt]
        \hline
        {\it TBabs} & $N_{\rm H}$  ($10^{20}$ cm$^{-2}$)   & \U{2.0}{2.8}{2.0} & $< 1.4$ & $<28$\\[2pt]
        \hline
        {\it bbodyrad}$_{\rm \,X}$ & $kT_{\rm bb}$ (keV) &   - & -    & \U{0.058}{0.059}{0.058}  \\[2pt]
        & $N_{\rm bb}$ \tnote{{\it{a}}} & - & - & \U{37.7}{\ast}{37.7} \\[2pt]
        \hline
        {\it comptt} & $kT_0$ (keV)  & - & - & [\U{0.058}{0.059}{0.058}]\tnote{{\it{b}}}\\[2pt]
        & $kT_e$ (keV)  & - & - & [100] \\[2pt]
        & $\tau$  & - &-  & \U{0.19}{0.15}{0.07} \\[2pt]
        & $N_{\rm c}$ ($10^{-7}$) & - & - &  \U{3.6}{\ast}{1.9}\\[2pt]
        \hline
        {\it diskir}&$kT_{\rm in}$ (keV)& \U{0.21}{0.02}{0.03} & \U{0.086}{0.03}{0.04}& - \\[2pt]
        & $\Gamma$ & \U{2.1}{\ast}{1.0}&\U{2.45}{0.17}{0.21} & -\\[2pt]
        &$kT_{\rm e}$ (keV) & [100]  & [100] & -\\[2pt]
        & $L_{\rm c}/L_{\rm d}$ & \U{0.066}{9.5}{0.041}& \U{2.0}{\ast}{0.6} & - \\[2pt]
        & $f_{\rm in}$ & [0.1] & [0.1] & - \\[2pt]
        & $r_{\rm irr}$ & [1.2] & [1.2] & - \\[2pt]
        & $f_{\rm out}$ $(10^{-3})$ & \U{3.6}{7.4}{3.1} &  \U{27}{\ast}{12} & - \\[2pt]
        & log($R_{\rm out}$) & \U{3.57}{0.13}{0.18} & \U{3.16}{0.30}{\ast} & - \\[2pt]
        &$K$\tnote{{\it{c}}}
        & \U{27.5}{35.4}{11.1}& \U{36.5}{89}{26.5}& -\\[2pt]
        \hline
        {\it bbodyrad}$_{\rm \,B}$ & $T_{\rm bb}$ (K) & [21,170] & [21,170] & \U{21,170}{3760}{2890}\\[2pt]
        & $N_{\rm bb}$ ($10^9$)\tnote{{\it{a}}} & [3.6] & [3.6] & \U{3.6}{2.3}{1.4} \\[2pt]
        \hline
        {\it bbodyrad}$_{\rm \,R}$ & $T_{\rm bb}$ (K) & [5550] & [5550] & \U{5550}{250}{290}\\[2pt]
        & $N_{\rm bb}$ ($10^9$)\tnote{{\it{a}}} & [3.4] & [3.4] & \U{3.4}{3.0}{1.7} \\[2pt]
        \hline
        {\it mekal} & $kT_{\rm mk}$ (keV) & [0.44] & [0.44] & \U{0.44}{0.14}{0.13}\\[2pt]
          & $N_{\rm mk}$ ($10^{-6}$) & [1.1] & [1.1] &  \U{1.1}{0.7}{0.5}\\[2pt]
        \hline
        & $f_{\rm X}$ ($10^{-14}$  erg cm$^{-2}$ s$^{-1}$)\tnote{{\it{d}}}  & \U{57}{2}{2}& \U{7.2}{0.4}{0.2} & \U{2.6}{0.2}{0.2} \\[2pt]
        & $L_{\rm X} \cos \theta$  ($10^{40}$ erg s$^{-1}$)\tnote{{\it{e}}} & \U{38.1}{6.0}{4.9} &\U{4.1}{0.3}{0.2}& \U{2.8}{0.4}{0.2} \\[2pt]        
        \hline
        & $f_{\rm O}$ ($10^{-14}$  erg cm$^{-2}$ s$^{-1}$)\tnote{{\it{f}}}  & \U{1.9}{0.1}{0.1}& \U{1.3}{0.1}{0.1} & \U{0.56}{0.04}{0.04} \\[2pt]
        & $f_{\rm B}$  ($10^{-14}$ erg cm$^{-2}$ s$^{-1}$)\tnote{{\it{g}}} & [0.37] & [0.37] & \U{0.37}{0.03}{0.01} \\[2pt]
        & $f_{\rm R}$  ($10^{-14}$ erg cm$^{-2}$ s$^{-1}$)\tnote{{\it{h}}} & [0.19] & [0.19] & \U{0.19}{0.02}{0.04} \\[2pt]
        \hline
        & $R_{\rm in} \, \sqrt{\cos \theta}$ ($10^8$ cm) & \U{57}{30}{13} & \U{66}{57}{31} & - \\[2pt]
        \hline
        & $\chi^2_{\nu}$ & 1.25 (88.9/71) & 0.89 (64.8/73) & 0.73 (61.7/85) \\[2pt]
        \hline
    \end{tabular}
        \begin{tablenotes}
      {\footnotesize{
      \item[{\it {a}}]{$N_{\rm bb} \equiv \left(R_{\rm bb}/{\rm{km}}\right)^2\,\left(10\,{\rm {kpc}}/d \right)^2$.}
      \item[{\it {b}}]{Seed temperature of the {\it comptt} model locked to the blackbody temperature.}
      \item[{\it {c}}]{$K \equiv \left(r_{\rm in}/{\rm{km}}\right)^2\,\left(10\,{\rm {kpc}}/d \right)^2 \, \cos \theta$.}
      \item[{\it {d}}] Observed flux in the 0.3--10 keV band (not including the {\it mekal} component).
      \item[{\it {e}}]{Unabsorbed luminosity in the 0.3--10 keV band (not including the {\it mekal} component); $L_{\rm X} \equiv (2\pi \, d^2/\cos \theta)\, f_{\rm X}$ for the 2010 (high) and 2012 (intermediate) states, and $L_{\rm X} \equiv 4\pi \, d^2\, f_{\rm X}$ for the 2013 (low) state.
      \item[{\it {f}}] Total observed flux at $\lambda > 912$ \AA.
      \item[{\it {g}}] Observed flux at $\lambda > 912$ \AA, attributed to a constant blue component.
      \item[{\it {h}}] Observed flux that is attributed to a constant red component.
      }
      }}
    \end{tablenotes}
  \end{threeparttable}
    \vspace{0.3cm}
    \label{tab_diskir}
\end{table*}


\begin{figure*}
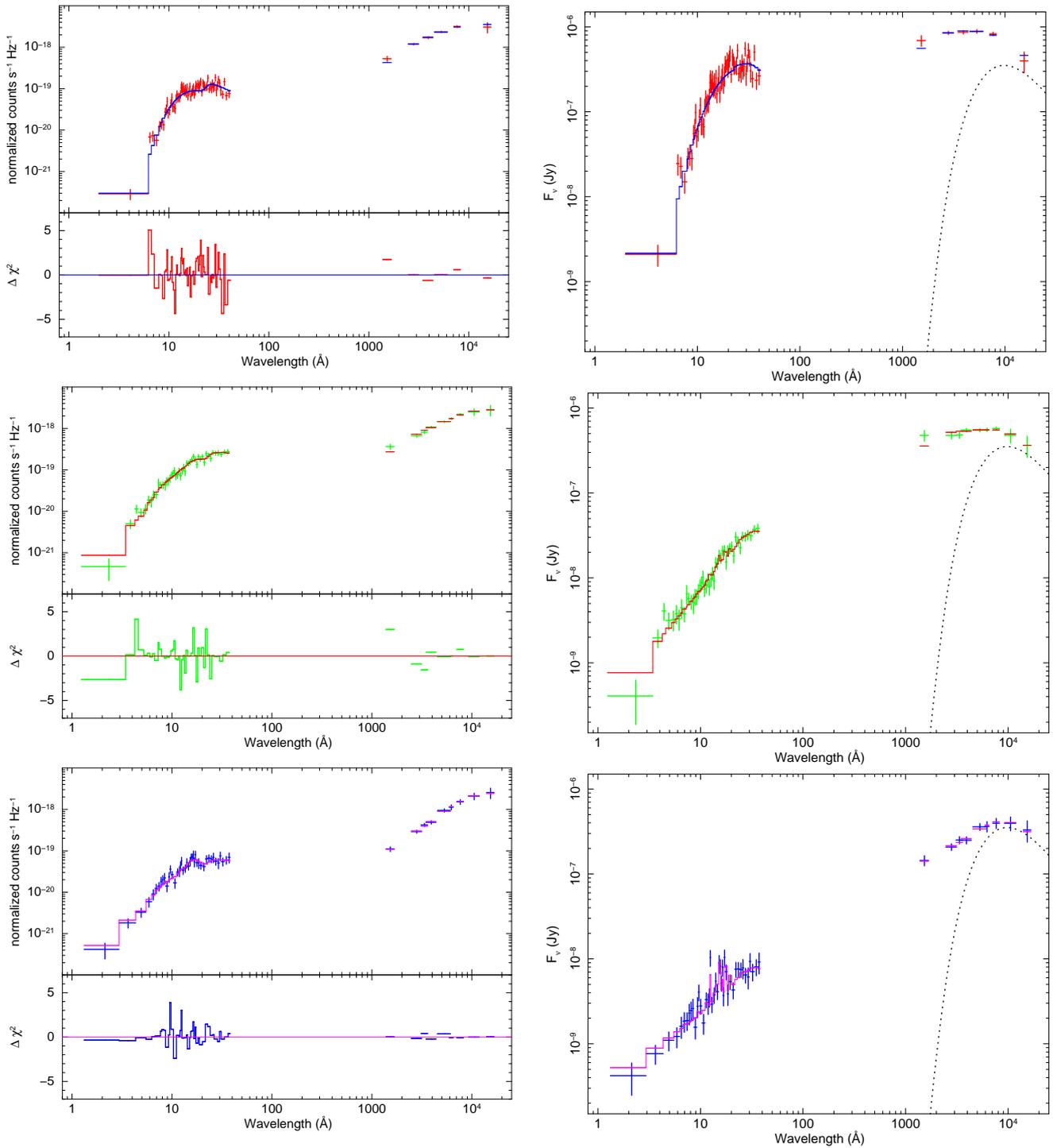

\hspace{-0.4cm}
\psfig{figure=new_2010_model1_plot1.eps,height=8.8cm,angle=270}
\psfig{figure=new_2010_model1_plot2.eps,height=8.8cm,angle=270}\\
\hspace{-0.3cm}
\psfig{figure=new_2012_model1_plot1.eps,height=8.8cm,angle=270}
\psfig{figure=new_2012_model1_plot2.eps,height=8.8cm,angle=270}\\
\hspace{-0.3cm}
\psfig{figure=new_2013_model1_plot1.eps,height=8.8cm,angle=270}
\psfig{figure=new_2013_model1_plot2.eps,height=8.8cm,angle=270}
\caption{Top row, left panel: spectral energy distribution and fit residuals for the broad-band spectrum of HLX-1 in the high/soft state (2010 {\it HST} data plus stacked {\it Swift} spectrum), fitted with the model summarized in Table 2 (Model 1). Top row, right panel: unfolded spectrum of the high/soft state; the dotted line represents the constant blackbody component (which we attribute to an old stellar population) in our model. Middle row, left panel: fitted spectrum and residuals in the intermediate state (2012 {\it HST} and {\it XMM-Newton} data). Middle row, right panel: unfolded spectrum in the intermediate state, with the constant blackbody component marked by a dotted line. Bottom row, left panel: fitted spectrum and residuals in the low/hard state (2013 {\it HST} and {\it XMM-Newton} data). Bottom row, right panel: unfolded spectrum in the low/hard state, with the blackbody component marked by a dotted line.}
\label{fig5}
\end{figure*}

\begin{figure*}
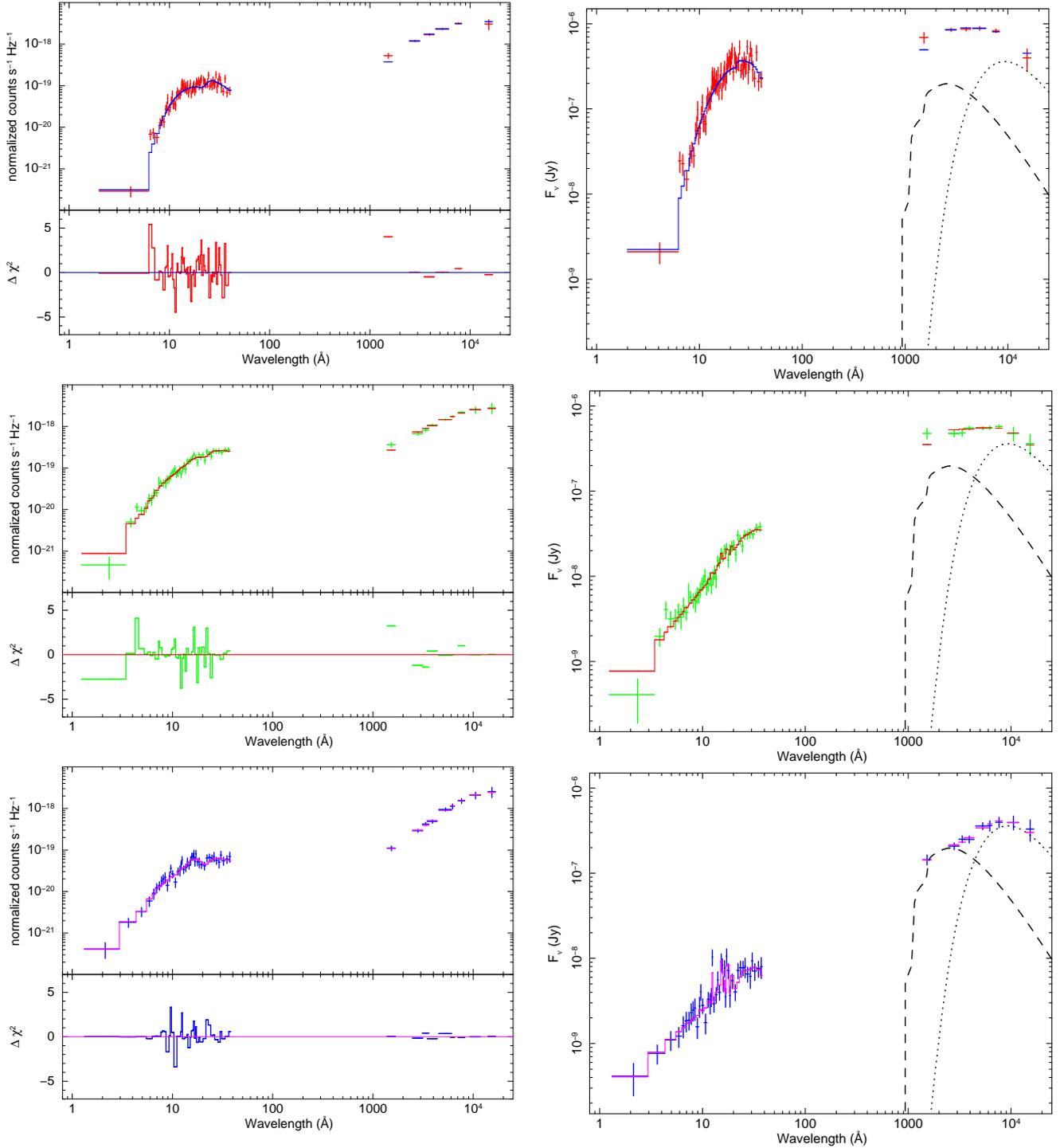

\hspace{-0.4cm}
\psfig{figure=new_2010_model2_plot1.eps,height=8.8cm,angle=270}
\psfig{figure=new_2010_model2_plot2.eps,height=8.8cm,angle=270}\\
\hspace{-0.3cm}
\psfig{figure=new_2012_model2_plot1.eps,height=8.8cm,angle=270}
\psfig{figure=new_2012_model2_plot2.eps,height=8.8cm,angle=270}\\
\hspace{-0.3cm}
\psfig{figure=new_2013_model2_plot1.eps,height=8.8cm,angle=270}
\psfig{figure=new_2013_model2_plot2.eps,height=8.8cm,angle=270}
\caption{Top row, left panel: spectral energy distribution and fit residuals for the broad-band spectrum of HLX-1 in the high/soft state (2010 {\it HST} data plus stacked {\it Swift} spectrum), fitted with the model summarized in Table 3 (Model 2). Top row, right panel: unfolded spectrum of the high/soft state; the dotted line is the redder blackbody component (which we attribute to an old stellar population), and the dashed line is the bluer blackbody component (which we associate with a younger stellar population) in our model. Middle row, left panel: fitted spectrum and residuals in the intermediate state (2012 {\it HST} and {\it XMM-Newton} data). Middle row, right panel: unfolded spectrum in the intermediate state, with the constant red and blue blackbody components marked by a dotted and a dashed line, respectively. Bottom row, left panel: fitted spectrum and residuals in the low/hard state (2013 {\it HST} and {\it XMM-Newton} data). Bottom row, right panel: unfolded spectrum in the low/hard state, with the red and blue blackbody components marked by a dotted and a dashed line, respectively.}
\label{fig6}
\end{figure*}

Given the small number of optical/UV datapoints, and therefore the small number of free parameters we can introduce, we tried fitting the broad-band data with two simple models: one in which all of the near-UV emission comes from the irradiated disk at all epochs (``Model 1"), and one in which all of the near-UV emission in the faintest epoch (2013) comes from a young stellar population, and the additional irradiation component only appears in 2010 and 2012 (``Model 2"). Those are clearly extreme cases: in reality, the contribution from the young stellar population may be somewhere between the two cases.

Let us start with Model 1, which consists of an irradiated disk ({\it diskir} in {\small XSPEC}: \citealt{gierlinski08,gierlinski09}) plus a constant, cool blackbody component ({\it bbodyrad}). 
The parameters of the blackbody component are determined from the 2013 observation, when it dominates the optical spectrum. These parameters are then kept fixed for our fits to the 2010 and 2012 {\it HST} data, dominated by the bluer component. We find (Figure 5, Table 2) that we can reproduce the spectral energy distribution at all three epochs. The moderately high fit residuals in the 2010 dataset ($\chi^2 \approx 1.25$) are mostly due to the X-ray part of the spectrum (the stacked {\it Swift}/XRT data), and to a near-UV excess in the F140LP filter. The latter is probably caused by the fact that this band was observed 10 days before the other {\it HST} bands, closer to outburst peak, and the near-UV is the most sensitive colour to the effect of X-ray irradiation. More detailed analyses and discussions of the 2010 {\it HST} and broadband data have already been presented elsewhere \citep{farrell12,soria12,mapelli13a,farrell14} and need not be repeated here. The new, interesting result of this work is that the irradiated disk plus blackbody model formally works also for the intermediate and low state observations. However, a very high reprocessing fraction $f_{\rm out} \approx 4\%$ is required in 2012 and 2013. 

In simple terms, the high value of $f_{\rm out}$ in the 2012 and 2013 fits is because the blue flux (modelled with irradiation) decreases more slowly than the soft X-ray flux in the three {\it HST} epochs used for our modelling, from 2010 to 2012 and 2013. In 2010, the observed 0.3--10 keV flux was $f_{\rm X} \approx 5.7 \times 10^{-13}$ erg cm$^{-2}$ s$^{-1}$, and the flux at $\lambda > 912$\AA\ was $f_{\rm O} \approx 2.4 \times 10^{-14}$ erg cm$^{-2}$ s$^{-1}$, of which $f_{\rm R} \approx 1.8 \times 10^{-15}$ erg cm$^{-2}$ s$^{-1}$ modelled as a constant red component. That gives a ratio $f_{\rm X}/(f_{\rm O}-f_{\rm R}) \approx 26$ between the short- and long-wavelength emission of the accretion flow. In 2012, $f_{\rm X} \approx 7.2 \times 10^{-14}$ erg cm$^{-2}$ s$^{-1}$, and $f_{\rm O} \approx 1.4 \times 10^{-14}$ erg cm$^{-2}$ s$^{-1}$, so that $f_{\rm X}/(f_{\rm O}-f_{\rm R}) \approx 6$. In 2013, $f_{\rm X} \approx 2.6 \times 10^{-14}$ erg cm$^{-2}$ s$^{-1}$, and $f_{\rm O} \approx 6.2 \times 10^{-15}$ erg cm$^{-2}$ s$^{-1}$, also for a ratio $f_{\rm X}/(f_{\rm O}-f_{\rm R}) \approx 6$.

At first sight, such high levels of optical reprocessing appear difficult to explain. They are an order of magnitude higher than predicted by standard thin-disk models \citep{dubus99,king97,dejong96,vrtilek90}, supported by observations of sub-Eddington Galactic X-ray binaries \citep{russell14,gierlinski09,hynes02}. In ultraluminous X-ray sources (ULXs), which in most cases are likely to be super-Eddington accretors, it is harder to distinguish between the emission from the donor star and from the irradiated disk \citep{heida14,gladstone13,grise12,tao12,tao11}; therefore, it is also more difficult to determine the disk reprocessing fraction. In one ULX where all optical emission was proved to be from the irradiated disk (M\,83 ULX), a reprocessing factor of $\approx$5 $\times 10^{-3}$ was inferred \citep{soria12b}; however, for other ULXs, broad-band emission models suggested reprocessing factors of a few $10^{-2}$ \citep{sutton14}. The geometric solid angle subtended by the disk is insufficient to explain such reprocessing factors only from direct X-ray illumination; however, if there is a strong outflow launched from the inner part of the disk, it was suggested \citep{sutton14,narayan17} that some of the X-ray photons emitted along the polar funnel may be scattered isotropically and contribute to the illumination of the outer disk. In this paper, we argue that HLX-1 is a sub-Eddington source (IMBH accretor), so in that sense it would be more appropriate to compare its reprocessing fraction with those of Galactic X-ray binaries. On the other hand, we will also argue (Section 5.3) that it has a strong wind, which would be consistent with the photon scattering scenario proposed for ULXs and with a high reprocessing factor. 

In fact, when we examine more carefully the physical meaning of $f_{\rm out}$ in 2012 and 2013, we propose other explanations for those high values. In 2012, optical and X-ray observations were not strictly simultaneous: the {\it XMM-Newton} observation happened 4 days after the {\it HST} observation. Those few days between the two observations are precisely the moment when HLX-1 started to switch from the high/soft to the low/hard state, with a drop in the {\it Swift}/XRT count rate by an order of magnitude (Figure 2, bottom panel). Thus, when the {\it HST} measurements were taken, the X-ray flux was almost certainly a few times higher than what was measured with {\it XMM-Newton} a few days later. We fitted the {\it XMM-Newton} and {\it HST} data together without accounting for the decrease in the X-ray flux: this means that we are almost certainly over-estimating by the same amount the true value of $f_{\rm out}$ needed to match the X-ray and UV portions of the SED. As for the high reprocessing fraction fitted to the 2013 data, we recall that our Model 1 represents the extreme case of no contribution from a young stellar population: in this sense, $f_{\rm out}$ here represents the hard upper limit to the reprocessing fraction in 2013. In the opposite case (Model 2), the same spectral energy distribution can be fitted with the other extreme case of $f_{\rm out} \rightarrow 0$.

Let us consider now Model 2 (Table 3 and Figure 6). We start from the 2013 dataset, which gives us the constraint on the stellar contribution (no optical/UV disk emission).
On the X-ray side, we replace the {\it diskir} model with a simple blackbody plus Comptonization model ({\it bbodyrad} $+$ {\it comptt}), which does not extend into the UV. On the optical/UV side, we fit the 2013 spectrum only with two blackbody components, one bluer and one redder, with no additional contribution from a disk. Having determined the old and young stellar contributions from the 2013 data (Table 3), we impose that the same two components are also present in the 2010 and 2012 spectra with the same (fixed) temperature and normalization, in addition to an irradiated disk component.  

In summary, we find that Model 1 and Model 2 are statistically equivalent in all 3 epochs (compare Table 2 and Table 3). We cannot tell the difference between the scenarios of near-UV emission from the disk only, or from a young stellar population plus an irradiated disk. However, we can use the models to calculate the age and mass of the old stellar populations, and to put useful upper limits to the young stellar population. 
In Model 1, with only a red component, the best-fitting blackbody temperature corresponds to a dereddened $V = 25.56 \pm 0.10$ mag in the Vega system, with $V-I = 1.03 \pm 0.15$ mag, $B-V = 0.82 \pm 0.15$ mag. At a distance modulus of 34.8 mag (92 Mpc), the absolute brightness is $M_V \approx -9.2$ mag. We ran simulations of star cluster evolution with {\small Starburst99} Version 7.0.1 \citep{leitherer99,leitherer14}, for instantaneous star formation and metallicity $Z = 0.008$. 
We found that those optical colours and luminosities are consistent either with an intermediate-age star cluster with mass $M_{\ast} \approx 2 \times 10^5 M_{\odot}$ and age of $\approx$800--900 Myr, or with an old star cluster, in particular one with a mass $M_{\ast} \approx 3 \times 10^6 M_{\odot}$ and age of $\approx$6--8 Gyrs (Figure 7).
We repeated the same analysis for Model 2, with two optical blackbody components (Table 3). The colder component has a dereddened $V = 25.41 \pm 0.10$ mag, $V-I = 0.93 \pm 0.15$ mag, $B-V = 0.76 \pm 0.15$ mag, similar to the red component in the first model. The hotter component has $V = 26.23 \pm 0.10$ mag, $V-I = -0.28 \pm 0.15$ mag, $B-V = -0.08 \pm 0.15$ mag. This is roughly consistent with the colours of a very young star cluster (age $\la 3$ Myr), with a mass $\la 10^4 M_{\odot}$, which can be taken as the firm upper limit to the young stellar component associated to HLX-1. More detailed spectral modelling, with proper spectral energy distributions for stellar populations of various ages and metallicities, in place of simple blackbody components, is left to follow-up work (C.~Maraston, priv.~comm.). 

The 2013 X-ray spectral fit is statistically improved (with F-test significance $> 99.9\%$) by the addition of a thermal-plasma component ({\it mekal} model in {\small XSPEC}), with solar abundance and fixed redshift $z = 0.0224$. We determined the temperature and normalization of the thermal plasma emission using the same strategy that we applied to the optical blackbody components: namely, we left those two {\it mekal} parameters free to vary in the 2013 spectrum, determined their best-fitting values, and then kept them frozen in the 2010 and 2012 spectra, assuming that the thermal plasma component does not vary on short timescales. (In the 2010 and 2012 X-ray spectra, the additional thermal plasma component also improves the fit but with $< 90\%$ significance, because of the comparatively stronger disk emission). We found a best-fitting {\it mekal} temperature of $\approx (0.5 \pm 0.1)$ keV, and a de-absorbed luminosity of $\approx (3.3 \pm 1.3) \times 10^{39}$ erg s$^{-1}$, almost identical in Model 1 and Model 2. This is also the same thermal-plasma temperature and luminosity that was found by \cite{servillat11} in an earlier {\it XMM-Newton} spectrum from 2010 May 14, also in the low/hard state (in-between the 2009 and 2010 outbursts). 
\cite{servillat11} used the higher spatial resolution of {\it Chandra}/ACIS-S to show (see their Sections 3.2--3.3) that the soft thermal component is likely to be diffuse emission from the bulge of ESO\,243-49 rather than from HLX-1. Such extended component cannot be resolved in {\it XMM-Newton} (the 30'' source extraction region for HLX-1 includes also the nuclear region and most of the galaxy). Thus, the thermal plasma emission seen in the {\it XMM-Newton} spectra of HLX-1 may have a different physical interpretation than the thermal plasma emission seen in several ULXs \citep{middleton15,sutton15,pinto16,urquhart16}. In the latter group of sources, the line residuals appear to come from massive outflows, directly associated with the accreting compact objects; they are interpreted as evidence that those sources are stellar-mass BHs or neutron stars accreting much above their Eddington limit. The lack of intrinsic X-ray thermal plasma features in the HLX-1 spectra, on the other hand, is consistent with the interpretation of this source as an IMBH accreting at an Eddington rate $\la$1. For these reasons, we did not include the flux and luminosity of the thermal plasma component in the values of $f_{\rm X}$ and $L_{\rm X}$ reported in Tables 2 and 3.


\begin{figure}
\hspace{-0.4cm}
\psfig{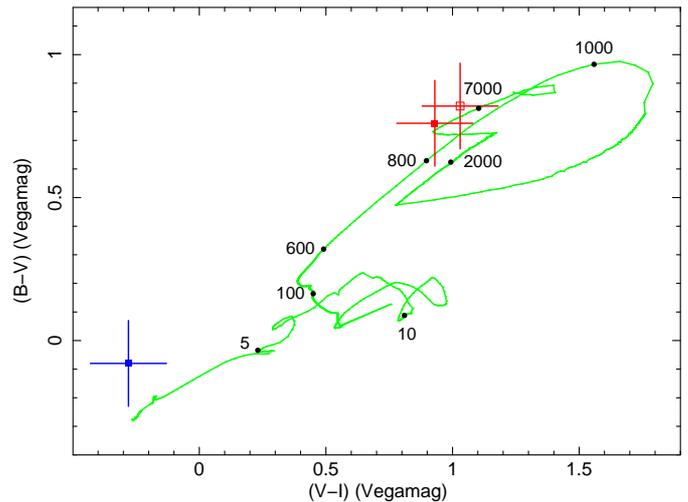}
\caption{Open red square: dereddened optical colours (Vegamag system) of the cooler blackbody component ($T \approx 5300$ K) in the optical spectrum of HLX-1, for Model 1. Filled red square: dereddened colours of the cooler component ($T \approx 5500$ K) in Model 2. Filled blue square: dereddened colours of the hotter blackbody component ($T \approx 21,000$ K) in Model 2. Green curve: predicted colours of a single-population star cluster as a function of age. A few characteristic ages in Myr have been labelled along the curve.}
\label{fig7}
\end{figure}


\section{What is the mass of the black hole?}


The normalization of the {\it diskir} model ({\i.e.}, the fitted inner radius of the disk) in the high/soft and intermediate state (Tables 2 and 3) is consistent between Model 1 and Model 2. 
By taking a weighted average of the four best-fitting values from the high and intermediate state (two from Model 1 and two from Model 2) we obtain:
\begin{equation}
R_{\rm in} \sqrt{\cos \theta} \approx 1.19 \, r_{\rm in} \sqrt{\cos \theta} \approx \left(49.3^{+12.4}_{-5.6}\right) \times 10^3 \ \ {\mathrm {km}},
\end{equation}
where the factor 1.19 takes into account the hardening factor and the inner-boundary condition, as discussed in \cite{kubota98}. (Notice that the four values used for this average are all consistent with each other within their errors.)
Our best-fitting inner-disk radius is also consistent with those previously published in the literature \citep{farrell14,farrell12,soria12,godet12,soria11,farrell09}. 
Instead, in the low/hard state, the size of the thermal seed component (disk-blackbody in the {\it diskir} model and blackbody in the {\it comptt} model) cannot be constrained: the data are also consistent with seed temperatures $\la 50$ eV (a range in which they can no longer be meaningfully constrained with the EPIC detectors), and a correspondingly larger normalization. The {\it diskir} model does suggest that $R_{\rm out}/R_{\rm in} \la 1400$ in the low/hard state (Table 2), a few times less than in the high state (when $R_{\rm out}/R_{\rm in} \approx 4000$--5000); if we assume that the outer disk radius remains approximately constant (as determined in Section 4.2), it means that the inner radius is moving further out (truncated disk), to a value $R_{\rm in} \ga 2 \times 10^5$ km, consistent with the canonical evolution of an accretion disk in BH transients at the end of an outburst.

Henceforth, we will use only the high-state and intermediate-state spectra for an estimate of the inner-disk radius and BH mass.
The fact that in several disk-dominated X-ray spectra, measured with different instruments over different outbursts, the inner disk radius is consistently found to be $R_{\rm in} \, \sqrt{\cos \theta} \approx 50,000$--$100,000$ km (see also Table 3 in \citealt{farrell14}) suggests that this parameter is physically meaningful, representing the innermost stable circular orbit $R_{\rm isco}$. We introduce a parameter $\alpha_{\rm s}$ to express the innermost stable orbit as function of the gravitational
radius, such that $R_{\rm in} = R_{\rm isco} \equiv \alpha_{\rm s} \, GM/c^2$, where $1 < \alpha_{\rm s} \le  6$, depending on the BH spin. In the framework of the standard disk model, the BH mass is then 
\begin{equation}
M \approx \frac{(3.3^{+0.9}_{-0.4})}{\alpha_{\rm s} \, \sqrt{\cos \theta}} \times 10^4 M_{\odot}.
\end{equation}
If we assume that the peak of each outburst corresponds to the Eddington luminosity (in the X-ray band), we obtain $M \approx 1 \times 10^4 M_{\odot}$ and $\alpha_{\rm s} \, \sqrt{\cos \theta} \approx 3.3$. However, there is no compelling reason for assuming Eddington-limited outbursts: many Galactic X-ray transients peak at only a fraction of Eddington \citep{remillard06,fender04}. 

Fortunately, the {\it XMM-Newton} spectrum from 2012 November 23 provides new information on a particularly interesting stage of the outburst cycle, and an additional independent constraint to the BH mass. In the model-independent hardness-intensity diagram, based on the observed EPIC-pn count rates, we note (Figure 8) that the position of the source appears to fall in between the high/soft and the low/hard state. Using Galactic X-ray binary terminology, we could say that the system was on the lower/descending branch of the so called ``Q diagram", which is often used to describe phenomenologically the evolution of stellar-mass BH transients \citep{fender04}. In fact, a full ``Q" cycle has never been observed for HLX-1, perhaps because the outburst rise in the hard state and the transition to the soft state occur very quickly; based on {\it Swift} data, the hardness-intensity diagram for HLX-1 shows two clear states but little evidence of the track between the two \citep{yan15}. Regardless of the actual shape of the outburst cycle, the 2012 data suggest that HLX-1 was in the middle of the (fairly rapid) transition from the soft to the hard state. This scenario is confirmed by the sharp decline in the {\it Swift}/XRT count rate that occurs right at the time of the 2012 observation (Figure 2, bottom panel), over a timescale of less than a week. Our spectral modelling shows (Tables 2 and 3) that disk and Comptonized (power-law) components contribute at comparable levels; the power-law photon index is still relatively soft ($\Gamma \approx 2.2$), while the peak disk temperature $kT_{\rm in} \approx 120$ eV is a factor of 2 below the temperature at outburst peak. 

Empirical evidence from Galactic X-ray transients shows that the soft-to-hard state transition at the end of an outburst occurs at a characteristic luminosity $L \approx L_{\rm X} \approx 0.01$--0.03 $L_{\rm Edd}$ \citep{maccarone03,kalemci13}. We determined a de-absorbed luminosity $L_{\rm X} = (4.1^{+0.3}_{-0.2} \times 10^{40})/\cos \theta$ erg s$^{-1}$ for the 2012 {\it XMM-Newton} data, assuming a disk-like emission geometry. Taking $L_{\rm Edd} \approx 1.3 \times 10^{38} (M/M_{\odot})$ erg s$^{-1}$, we obtain:
\begin{equation}
    M \approx \frac{1.6^{+1.7}_{-0.6}}{\cos \theta}\, 10^4 \, M_{\odot}.
\end{equation}
Combining Equations (2) and (3), we derive: 
\begin{equation}
\alpha_{\rm s} \approx \left(2.1^{+1.3}_{-1.1}\right) \, \sqrt{\cos \theta}.
\end{equation}
Since $\cos \theta \le 1$, $R_{\rm isco} \la 3.4 GM/c^2$, which implies that the BH spin parameter $a/M \ga 0.7$. We do not have any direct measurements of $\cos \theta$; however, the low absorbing column density required by the X-ray spectral fits suggests that HLX-1 is more likely not seen at very high inclination. This is also consistent with the viewing angle $\theta \approx 30^{\circ}$ proposed by \cite{cseh15} to model the Doppler boosting of the radio jet in the low/hard state. The central value of our estimate (corresponding to a soft-to-hard transition luminosity of 0.02$L_{\rm Edd}$) is $\alpha_{\rm s} \approx 2.1$ for $\theta \approx 0^{\circ}$, $\alpha_{\rm s} \approx 2.0$ for $\theta \approx 30^{\circ}$, $\alpha_{\rm s} \approx 1.8$ for $\theta \approx 45^{\circ}$.
This implies a spin parameter $a/M \approx 0.9$, almost independent of $\theta$, for $\theta \la 45^{\circ}$.

In conclusion, from the transition luminosity and the fitted inner disk radius, and for moderately face-on viewing angles, we estimate a BH mass $M \approx (2\pm 1) \times 10^{4} M_{\odot}$, a BH spin parameter $\ga 0.7$ for any viewing angle, and $\approx 0.9$ for face-on angles, a peak outburst luminosity $L_{\rm X} \approx (0.3 \pm 0.15)\,L_{\rm Edd}$. Our best-fitting values of mass and spin are very similar to those inferred by \cite{davis11} (see in particular their Fig.~5) from {\it Swift} data, in the case of $\theta \approx 0$. Our best-fitting BH mass is also similar to the value found by \citep{godet12}, although the latter result was based on slim-disk models, which may not be relevant if the peak luminosity is always sub-Eddington. Note that our value of the peak outburst luminosity is about half the value usually reported in the literature ({\it e.g.}, \citealt{farrell09,farrell12,yan15}), because we have defined luminosities as $(2\pi/\cos \theta)$ times the flux (as more suitable to disk-dominated emission) rather than $4\pi$ times the flux. 

If we assume instead a perfectly isotropic emission during the 2012 intermediate state, the luminosity becomes $L_{\rm X} = (8.2^{+0.6}_{-0.4} \times 10^{40})$ erg s$^{-1}$ $\approx 0.01$--0.03 $L_{\rm Edd}$. Repeating the same derivation, we obtain $M \approx (3.3^{+3.4}_{-1.1}) \times 10^4 M_{\odot}$, and $\alpha_{\rm s} \, \sqrt{\cos \theta} =  1.1^{+0.7}_{-1.1}$. For a moderately face-on view, this implies an even higher BH spin, close to an extreme Kerr BH. The peak outburst luminosity would be $L_{\rm X} \approx (0.15 \pm 0.08)\,\cos \theta \, L_{\rm Edd}$.

\section{How large is the disk?}

Following \cite{soria13a}, we estimate the outer radius $R_{\rm out}$ of the accretion disk in two independent ways. The first method is to assume that the X-ray luminosity during the initial phase of outburst decline has an exponential-decay timescale of order of the viscous timescale at the outer edge of the disk. This can be understood in the framework of thin accretion disk models \citep[{\it e.g.},][]{frank02}; it also has an empirical analogy with the temporal brightness variations of Galactic X-ray binaries during their outbursts \citep{king98}. The second method assumes that the optical emission comes mostly from the irradiated disk (as discussed in Section 2.3): the characteristic size of the thermal emitter is derived from the observed optical luminosity and temperature. 

\subsection{Disk size from X-ray outburst properties}

During the 2009, 2010 and 2011 outbursts, the X-ray emission shows characteristic exponential-decay timescales $\tau \approx (3$--$6) \times 10^6$ s \cite{soria13a}. For the 2012, 2013 and 2015 outbursts, the $e$-folding decay timescales are shorter, $\tau \approx 2 \times 10^6$ s \citep{yan15}, although this value is possibly affected by the earlier transition from the exponential to the linear decay phase (Fig.~6 in \citealt{yan15}). Such differences between outbursts are not crucial for our attempt to provide an order-of-magnitude estimate of the disk size. 
Exponential decay timescales of a few weeks are near the upper limit but still within the range of those observed in Galactic X-ray binary transients \citep{yanyu15,chen97}, despite the much higher BH mass proposed by the IMBH scenario.  

During the exponential decline of an outburst, when the rate of mass depletion in the disk (via accretion and/or outflows) is much higher than any ongoing mass transfer from the companion star, the decay timescale is related to the outer disk radius by the relation \citep{frank02}:
\begin{equation}
\tau \approx R^2_{\rm out}/(3\nu),
\end{equation}
where $\nu$ is the kinematic viscosity. In the Shakura-Sunyaev model \citep{ss73}, $\nu = \alpha c_{\rm s} H$, where $\alpha \sim 0.1$--$0.5$ is a scaling parameter (not to be confused with the BH spin parameter used in Section 3), $c_{\rm s}$ is the mid-plane sound speed in the outer disk and $H$ is the disk scale-height. In the simplest empirical approximation, we can take $\alpha (H/R) \sim 0.01$ \citep{shahbaz98}. Then, $R_{\rm out} \sim 0.03 c_{\rm s} \tau$. For HLX-1, we expect characteristic mid-plane temperatures $T \sim$ a few $10^5$ K in the outer disk \citep{frank02}; hence, $c_{\rm s} \sim 40$--60 km s$^{-1}$ and $R_{\rm out} \sim 10^{12}$ cm. Using a more accurate expression of $\nu$ from the Shakura-Sunyaev disk solutions with Kramer's opacity \citep{ss73,frank02}, for the observed X-ray properties of HLX-1, we re-obtain a characteristic disk radius $\la 10^{12}$ cm (Equation 10 in \citealt{soria13a}) for the earlier outbursts. For the timescales measured in the most recent outbursts, a more likely estimate of the characteristic radius is $R_{\rm out} \sim$ a few $10^{11}$ cm. 

 When the disk is completely ionized (a condition expected to be easily satisfied in HLX-1: \citealt{lasota11}), the mass content in the disk declines exponentially on the viscous timescale \citep{king98} after the peak of the outburst. This leads to a relation \citep{shahbaz98} between outer disk radius and peak luminosity $L_{\rm peak}$:
\begin{equation}
  R_{\rm out} \approx 1.0 \times 10^{12} \left(\frac{\eta}{0.1}\right)^{-1/3}  
  \left(\frac{L_{\rm peak}}{10^{42} {\rm erg~s}^{-1}}\right)^{1/3}  
  \left(\frac{\tau}{10^6 {\rm s}}\right)^{1/3} \ {\rm cm},
\end{equation} 
where $\eta \equiv L/(\dot{m}c^2)$ is the radiative efficiency. Again, for the observed decline timescales and peak luminosity, and efficiency $\eta \approx 0.3$ (suitable to a fast-spinning BH), we obtain that $R_{\rm out} \approx 10^{12}$ cm, consistent with our previous estimate. 

Another (not entirely independent) estimate of disk size is based on the total fluence or energy released during an outburst. For HLX-1, the energy emitted as X-rays (which is a sufficiently good approximation to the total energy, for our order-of-magnitude estimate) declined from $\approx$10$^{49}$ erg to $\approx$4 $\times 10^{48}$ erg over the six recorded outbursts (Table 3 in \citealt{yan15}), assuming isotropic emission, or a factor of 2 lower if we are looking at a face-on disk geometry. For a standard BH accretion efficiency $\eta \approx 0.1$, this energy corresponds to accreted masses between $\approx$10$^{29}$ g and $\approx$4 $\times 10^{28}$ g, in the earliest and latest outbursts, respectively. These values may be reduced by up to a factor of 2 in case of a face-one disk view, and another factor of 3 for a fast-spinning BH with $a/M \approx 0.9$. Considering the uncertainties, we can say that $10^{29}\,(0.1/\eta)$ g is a conservative upper limit to the mass accreted in each outburst.
Surface densities profiles $\Sigma(R)$ for irradiated disks around a 10-$M_{\odot}$ BH, for various accretion rates, were calculated by \cite{dubus99}. The main effect of irradiation is to keep the outer disk in the hot state; in that regime, the surface density is well approximated by the Shakura-Sunyaev solution
\begin{equation}
\Sigma \approx 5.2 \times 10^3 \, \alpha^{-4/5} \, \dot{M}_{21}^{7/10} \,
M_4^{1/4} \, R_{12}^{-3/4} \ {\mathrm {g~cm}}^{-2}    
\end{equation}
\citep{ss73,frank02}, where $M_4$ is the BH mass in units of $10^4 M_{\odot}$ and $\dot{M}_{21}$ the accretion rate in units of $10^{21}$ g s$^{-1}$ (suitable order of magnitude for the long-term-average accretion rate in HLX-1). The total mass $M_{\rm tot}$ in the disk up to a radius $R$ is given by 
\begin{eqnarray}
    M_{\rm tot}(R) &\approx& 2 \pi \int_{0}^{R} \Sigma \, R \, dR \nonumber \\
    &\approx& 2.6 \times 10^{28} \, \alpha^{-4/5} \, \dot{M}_{21}^{7/10} \, M_4^{1/4} \, R_{12}^{5/4}  \ {\mathrm g}.
\end{eqnarray}
Since the accreted mass during an outburst is $\la 10^{29} \,(0.1/\eta)$ g, and taking $\alpha \la 0.3$, $M_4 \approx 2$, from Equation (8) we conclude that a typical outburst can empty a disk (or a substantial fraction of the disk) of radius $R \la 10^{12}$ cm. If we also account for the factor-of-two reduction in the mass accreted in each outburst (in case of a face-on disk view), the disk size (or at least the portion of the disk) that can be emptied in each outburst is only $R \la 5 \times 10^{11}$ cm.

A radius $\sim$ a few $10^{11}$--$10^{12}$ cm is within the range of disk sizes in stellar-mass Galactic BH X-ray binaries (Fig.~1 in \citealt{remillard06}); for example, it would be smaller than the disk in GRS 1915$+$105 ($R_{\rm out} \approx 5 \times 10^{12}$ cm: \citealt{rau03}), similar to the disk radius in V404 Cyg ($R_{\rm out} \approx 9 \times 10^{11}$ cm: \citealt{munoz16}) and slightly larger than the disk radii inferred for GRO\,J1655$-$40, GRO\,1744$-$28, GX\,339-4 \citep{shahbaz98,homan05}. However, given the large difference in BH masses and therefore in the inner disk radius $R_{\rm in}$, for HLX-1 this would imply $R_{\rm out} \sim 100 R_{\rm in}$ while for Galactic stellar-mass BHs, $R_{\rm out} \sim 10^5 R_{\rm in}$.




\subsection{Disc size from optical continuum fitting}

As discussed in Section 2.3, the optical/UV continuum emission in the high and intermediate state must come mostly from an irradiated disk, with the addition of a red excess. Our {\it diskir} modelling of the 2010 {\it HST} data together with stacked high/soft {\it Swift} data (Tables 2 and 3) suggests an outer disk radius $R_{\rm out} \, \sqrt{\cos \theta} = (2.2^{+0.6}_{-0.4}) \times 10^{13}$ cm. This is consistent with the outer radii estimated by \cite{farrell12} and \cite{soria12}, and is a factor of 3 smaller than the value fitted by \citep{farrell14}; a value $R_{\rm out} \, \sqrt{\cos \theta} \approx 3 \times 10^{13}$ cm was estimated by \citep{mapelli13a} with the irradiated disk model of \cite{patruno08}. 
The main reason why our optical radius estimate is on the lower side of the measurement distribution is that we have attributed all the red and near-IR emission in the low-state 2013 {\it HST} dataset to a stellar component rather than the disk (Section 3.2), and have then kept this model component constant in the spectra for the the intermediate and high states. This reduces the outer radius of the disk required to fit the optical data.

The effective temperature at the outer edge of the disk (where irradiation dominates over viscous heating) can be estimated from the relation  
$\sigma [T(R_{\rm out})]^4 \approx \left(f_{\rm out} \, L_{\rm X}\right)/\left(4\pi R_{\rm out}^2\right)$, where $f_{\rm out}$ is the reprocessing fraction. From our {\it diskir} fit for the 2010 {\it HST} dataset, we infer $T(R_{\rm out}) \approx 20,000$ K. An alternative way of estimating the outer disk temperature is to fit the 2010 {\it HST} datapoints with a single blackbody component, because the irradiated disk emission is dominated by the blackbody emission from the largest annulus. We verified that this method also suggests an outer disk temperature of $\approx$20,000 K (for line-of-sight extinction).





In conclusion, the disk radius derived from optical photometry is at least $\approx$20 times, and more likely 40 times, larger than the radius derived from the X-ray lightcurve. The two values cannot be reconciled even by taking into account systematic uncertainties in both methods. While the outburst timescale is within the range of Galactic X-ray binaries, the peak optical brightness of the disk ($M_V \approx -11$ mag) is several magnitudes brighter; the most luminous irradiated disks in Galactic X-ray binaries reach only $M_V \sim -5$ mag \citep{vanp94}. In Section 5, we will discuss two scenarios that address this discrepancy: either by attributing the optical emission to a circumbinary (CB) disk, or by confining the cycle of X-ray outbursts to the innermost rings of the disk.

\begin{figure}
\hspace{-0.4cm}
\psfig{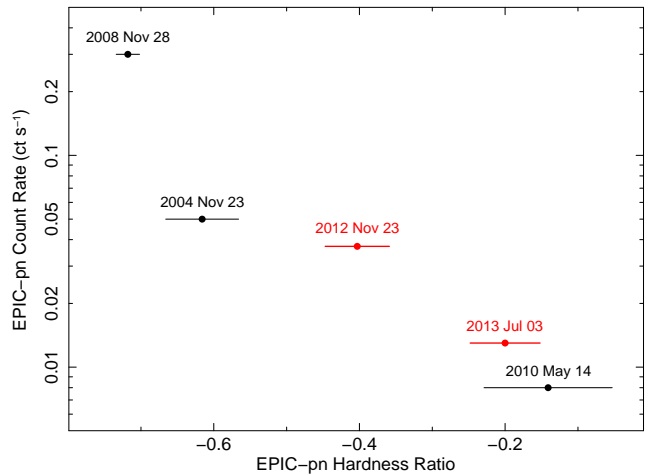}
\caption{Evolution of HLX-1 on the hardness-intensity plane, based on the observed EPIC-pn count rates in various observations. The hardness ratio (X axis) is defined as the ratio between the pn count rates in the 1--10 keV and 0.3--1 keV bands. The intensity (Y axis) is the pn count rate in the 0.3--10 keV band. Red datapoints correspond to the {\it XMM-Newton} observations analyzed in this work; black datapoints are from \citep{servillat11}. Observation dates are labelled.}
\label{fig8}
\end{figure}

\subsection{Implications for the system parameters}


In the absence of dynamical measurements for HLX-1, determining the outer disk radius provides an important constraint on the binary period and orbital separation, and therefore also on the type of donor star and mass transfer mechanism. Analytical models and numerical simulations suggest that for $q \equiv M_2/M_1 \ll 1$ (where $M_1$ is the BH mass and $M_2$ the donor star masss), the disk can extend to the tidal truncation radius $R_{\rm T}$: \begin{equation}
    R_{\rm out} \approx R_{\rm T} \approx 0.60\, a/(1+q),
\end{equation}
\citep{warner95,artymowicz94,whitehurst88,paczynski77,papaloizou77}. For the range of mass ratios of interest here ($q \sim 10^{-5}$--$10^{-3}$), the radius of the donor star (assumed to be filling its Roche lobe) is well approximated by 
\begin{equation}
    R_{2} \approx 0.49\, aq^{1/3}, 
\end{equation}
\citep{eggleton83,paczynski71}. Note also that for $q \ll 1$, the companion star fills its Roche lobe at a binary separation $a \approx 2 R_2 q^{-1/3}$, and gets tidally disrupted at a binary separation $a_{\rm td} \approx R_2 q^{-1/3}$ \citep{rees88}.
The binary separation is related to the binary period via Kepler's third law:
\begin{eqnarray}
    a &=& 1.50 \times 10^{13} \, (M_1 + M_2)^{1/3}\, P_{\rm {yr}}^{2/3} \ \ {\rm{cm}} \nonumber \\
    &=& 0.63 \times 10^{13} \, (M_1 + M_2)_4^{1/3}\, P_{\rm {d}}^{2/3} \ \ {\rm{cm}},
\end{eqnarray}
where the masses are in units of $M_{\odot}$ or $10^4 M_{\odot}$, respectively.

Here, we will briefly illustrate the characteristic system parameters corresponding to the two alternative cases of $R_{\rm out} \la 10^{12}$ cm and $R_{\rm out} \approx 2.5 \times 10^{13}$ cm (two representative values for the scenarios discussed in Section 4.1 and 4.2). We will also assume a BH mass of $\approx$$(2\pm1) \times 10^4 M_{\odot}$.

If it corresponds to the outer disk radius, the characteristic size derived from the X-ray lightcurve implies a binary separation $a \la 1.7 \times 10^{12}$ cm (Equation 9), and a binary period $P \la 2.3$ hr. The Roche lobe radius of the secondary is $R_2 \la 2.3 \times 10^{10}$ cm $\approx 0.33 R_{\odot}$ for $q = 2 \times 10^{-5}$, or $R_2 \la 3.7 \times 10^{10}$ cm $\approx 0.53 R_{\odot}$ for $q = 10^{-4}$.
Thus, the inferred constraint on the secondary Roche lobe can only be satisfied by an M-dwarf main-sequence donor star with $M_2 \la 0.4 M_{\odot}$ \citep{boyajian12}. However, it is also possible that the donor star has a radius much lower than its initial radius on the main sequence because it has already lost most of its outer envelope as a result of the intense mass transfer.  For a Roche-lobe-filling secondary star, and $0 < q \la 10^{-3}$, the period-density relation is well approximated by $P\rho^{1/2} \approx 0.41$ \citep{eggleton83}, where $P$ is in days and $\rho$ in g cm$^{-3}$. In our compact scenario for HLX-1, a period of $\approx$2.3 hours corresponds to a density $\rho \approx 18$ g cm$^{-3}$.  This is consistent with the density of an M3-M4 main-sequence star \citep{boyajian12}.

As discussed in Section 4.2, a small disk of radius $R_{\rm out} \la 10^{12}$ cm inside the BH Roche lobe cannot explain the bright optical emission. Therefore, we propose that the latter component comes from an irradiated CB disk. Smoothed particle hydrodynamics models show \citep{artymowicz94} that the inner boundary of a CB disk is truncated by tidal forces at a radius $R_{\rm CB,in}$ varying between $\approx$1.8$a$ and $\approx$3$a$ depending on orbital eccentricity. The difference is not significant for our case, because the emission dip expected from this gap falls in the far-UV band, where we do not have any measurements. Chris Copperwheat (priv.comm.) used his code \citep{copperwheat05,copperwheat07} to simulate irradiated disk spectra with or without a gap, and we verified that the difference is undetectable with the available X-ray and optical datapoints.
The outer radius of the irradiation-dominated CB disk would have to extend to $R_{\rm CB,out} \ga 15a \approx 2.5 \times 10^{13}$ cm to reproduce the optical continuum; its continuum emission  spectrum would look essentially identical to the spectrum of an irradiation-dominated accretion disk of the same size ($T(R) \propto R^{-1/2}$). See also \cite{farris14,farris15,yanlu15,dorazio16,dorazio13,roedig14,artymowicz96} for further modelling of CB disk structure and emission.

Observational evidence for mid-infrared emission from a dusty CB disk illuminated by the compact X-ray source has been found in several X-ray binaries, such as the Galactic BH candidates GRS\,1915$+$105 \citep{rahoui10}, A0620$-$00 and XTE\,J1118$+$480 \citep{muno06}. Among ULXs, a CB disk was proposed for Holmberg IX X-1 \citep{dudik16} and Holmberg II X-1 \citep{lau17}. A CB disk fed by outflows through the L2 point is also likely to be present in SS\,433, based on optical spectroscopic studies of its stationary H$\alpha$, Br$\gamma$, Paschen, and He\,I emission lines \citep{blundell08,perez09,perez10,bowler13}. From the rotational velocity shift of the red and blue line components, the characteristic emitting radius of the SS\,433 CB disk is $\sim$10$^{13}$ cm \citep{bowler13}. We have argued that in the CB disk scenario for HLX-1, its characteristic radius would also be $\sim$10$^{13}$ cm; however, its irradiating luminosity at outburst peak would be much higher than in SS\,433, and would keep the CB disk in the hot, optically-thick state with significant contributions to the UV continuum emission.

Let us now assume instead that the outer accretion disk radius is $R_{\rm out} \approx 2.5 \times 10^{13}$ cm, based on the optical continuum measurements. Following the same argument, for $q \ll 1$ and $M \approx 2 \times 10^4 M_{\odot}$, this disk radius implies a binary separation $a \approx 4 \times 10^{13}$ cm, and a binary period $P \approx 12$ d. The Roche lobe radius of the secondary is $R_2 \approx 5.5 \times 10^{11}$ cm $\approx 8 R_{\odot}$ for $q = 2 \times 10^{-5}$, or $R_2 \approx 9.5 \times 10^{11}$ cm $\approx 14 R_{\odot}$ for $q = 10^{-4}$, or $R_2 \approx 2.0 \times 10^{12}$ cm $\approx 29 R_{\odot}$ for $q = 10^{-3}$. From the period-density relation for $0 < q \la 10^{-3}$ we obtain a mean density $\rho \approx 1.2 \times 10^{-3}$ g cm$^{-3}$. This is consistent with several types of evolved stars ({\it e.g.}, blue supergiants, Hertzsprung gap stars, red giants). Massive stars in the last stages of stellar evolution are much more likely sources of continuously high mass-transfer rate (for timescales of a few $10^5$ yr) than a main-sequence M star. The only circumstance in which an M star can provide the required mass transfer rate is if it is currently undergoing tidal stripping of its envelope. 
Therefore, we conclude that the large disk scenario is much more consistent with the observations. This scenario would also imply that HLX-1 is fuelled in a way similar to ULXs, despite the likely large difference in the mass of the accretor. For completeness, in Section 5 we will outline possible outburst mechanisms both for the case of a smaller and of a larger orbital separation, keeping in mind that the latter case is more likely to apply to HLX-1.





\section{What is the outburst mechanism?}

\subsection{Summary of HLX-1 outburst models proposed in the literature}

The fast outburst rise and subsequent exponential decay, and the harder power-law spectrum in the low states, 
can be qualitatively explained with the disk instability model usually applied to X-ray binaries \citep{lasota08a,lasota08b,lasota01,dubus01,lasota00,menou00,dubus99,hameury98,hameury97}.   
In the model, 
the accretion disk extends to the innermost stable circular orbit 
$R_{\rm isco}$ when in the high state. 
Instead, in the low state, the disk is truncated at $R_{\rm tr} (\gg R_{\rm isco}$), within which the hot inflow is 
radiatively inefficient and geometrically thick.  
Moreover, in the low state, all disk annuli are on the cold branch 
  of the density-temperature S-curve 
  (characterised by neutral hydrogen and low viscosity), 
  with inflowing matter piling gradually 
  till the surface density exceeds the critical density.  
This occurs at $R$ slightly larger than $R_{\rm tr}$, 
  from where a heating front propagates outwards 
  (an inside-out outburst).  
The increased viscosity and accretion rate cause the inner disk to extend inwards 
  towards $R_{\rm isco}$, 
  and the system enters a high state, 
  where the entire disk is on the hot branch of the S-curve.  
When the outburst subsides and the accretion rate diminishes, the outer disk cools down and the inner disk evaporates, creating a radiatively inefficient, optically thin, geometrically thick inflow.

However, it was soon realized that this model does not work well for HLX-1 \citep{lasota11}. 
The observed X-ray luminosity of 
  $L_{\rm X} \approx 2 - 4 \times 10^{40}\;\!{\rm erg\;\!s}^{-1}$ 
  during the low/hard state \citep{servillat11,yan15}  
  requires an accretion rate larger than a few $10^{20}\;\!{\rm g\;\!s}^{-1}$  
  in the hot, radiatively inefficient inner region, 
  and hence $\dot{M}(R_{\rm tr}) \ga$ a few $10^{20}\;\! {\rm g\;\! s}^{-1}$. 
This implies that starting from a low-density disk in the cold state, 
  the transition radius at which the disk  
  becomes hot is 
  $R_{\rm tr} \approx 2 \times 10^{13}\;\! {\rm cm}$ \citep{dubus01,lasota08a}. 
If the inner disk was truncated at this large radius, it would take $\sim$10$^2$ yr to refill before it extends again to $R_{\rm isco}$.  
The high state would then be expected to last also for hundreds of years, 
  which is contradicted by the observed rapid recurrence times and short ourbursts. 
The disk instability model predicts $\dot{M}(R) \propto R^{2.65}$ 
  in the low state \citep{lasota01}. 
If we impose a smaller truncation radius in the low state, 
  $R_{\rm tr} \la 10^{12}\;\! {\rm cm}$, 
  to reconcile the refill time with the observed timescales, 
  and $\dot{M}(R_{\rm {tr}}) \approx$ a few 
  $10^{20}$ g s$^{-1}$,  
  we obtain  
  $\dot{M}(R_{\rm out}) \sim 10^{24}\;\!{\rm g\;\! s}^{-1} 
  \sim 10^{-2} M_{\odot}\;\! {\rm yr}^{-1}$, 
  a value completely unphysical for a donor star. 
Moreover, 
  the ratio between inter-outburst and outburst luminosity in HLX-1 
  is $\approx 3 \times 10^{-2}$, 
  about five to six orders of magnitude 
  higher than the ratios observed in transient BH X-ray binaries 
  and predicted by the disk instability model.
In summary, despite certain qualitative similarities 
  in the temporal and spectral properties 
  of HLX-1 and BH low-mass X-ray binaries, 
  the disk instability model fails to account for 
  the timescales of HLX-1's outburst duty cycle 
  and low/hard state luminosity 
  of $L_{\rm X} \sim 10^{40}\;\! {\rm erg\;\! s}^{-1}$.

One may explain the 
  repeat of the first few outbursts at almost the same time interval 
  of $\approx 1$~year 
  as successive bursts of mass transfer 
  from a donor star each time it passes through periastron of a very eccentric orbit \citep{lasota11}. 
Taking the binary period as 1 year, 
  the fitted BH mass of $\sim$10$^4 M_{\odot}$ 
  then sets the orbital semimajor axis $a \approx 3 \times 10^{14}\;\!{\rm cm}$ 
  ($\approx 20$~au).  
On the other hand, 
  the small disk size required by the fast rise and decay (radius $\la 10^{12}$ cm) 
  constrains the distance at the periastron, and hence the orbital eccentricity. 
The best solutions correspond 
  to extremely high elliptical orbits (almost reaching the parabolic limit) 
  \citep{lasota11,soria13a}.  
Such systems would be short-lived and/or unstable, as the entire envelope of the donor star would be 
  completely stripped after a few periastron passages. 
A less extreme alternative was proposed \citep{miller14}, 
  with an orbital eccentricity of ``only'' $\approx 0.9$, 
  where the mass transfer occurs via wind capture, 
  resulting in a lower specific angular momentum for the inflowing gas 
  and hence a shorter accretion timescale. 
This scenario still cannot avoid the donor star to be tidally stripped. 

Crucially, the recurrence time has increased substantially in the latest outbursts, up to 470 days between the   
 rise of the last two outbursts in 2013--2015 \citep{yan15},  
 and it has been a couple of years since the start of the last outburst in 2015 January. 
The periastron-passage scenario has to account for this development. 
Change in the orbital period can happen in highly eccentric, 
  extreme mass-ratio systems 
  undergoing mass loss and/or mass redistribution.  
The study of \cite{godet14} has shown that   
  a donor star captured by an IMBH 
  on an orbit with $e \ga 0.9998$ and periastron distance between $\approx$2 and 3
  times the tidal disruption radius 
  could reach a minimum period of about 1 year after a few dozen orbits.  
The star would lose a few percent of its mass at each passage, 
  leading to a rapid increase in the orbital period,  
  and eventually what is left of the star 
  becomes unbound after only 2 or 3 more orbits. 
This is consistent with the observations 
  and perhaps explains why there has been no outburst since 2015.
Smoothed particle hydrodynamics simulations of a similar scenario (but with different assumptions on orbital eccentricity, mass loss rates and viscosity) by \cite{vanderhelm16} showed that more moderate eccentricities such as $e=0.7$ or even $e=0.95$ are not sufficient to reproduce the rapid change in the recurrence timescale.

An important feature of all models based on extremely high eccentricity and tidal stripping is that the accretion disk must be small, with a maximum size comparable to the periastron distance, in agreement with the X-ray timescales rather than with the optically-measured size. That means that such models require 
an additional optically thick, irradiated gas region to account for the optical/UV continuum. For a similar reason, we have suggested the presence of a CB disk in the case of a compact system (Section 4.3). However, the CB disk scenario would itself have serious troubles, in the case of a highly eccentric orbit. In that situation, the apastron distance would determine the size of the gap between inner disk and the CB disk; therefore, the inner radius (hottest part) of the CB disk would have to be located at a distance $\ga 3 \times 10^{14}$ cm, if the semimajor axis of the orbit is large enough to correspond to a binary period of $\approx$ 1 yr (from Kepler's law). At those distances, the irradiated CB disk would be too cold to produce the observed UV/optical emission. Thus, a highly eccentric orbit would require an alternative, more complicated distribution of irradiated gas closer to the BH, possibly a transient combination of accretion streams, tidal streams and outflows around the stripped donor star.
The high-eccentricity, periastron-passage mass transfer scenario also implies that our chance detection of HLX-1 as a hyperluminous X-ray source is an one-off event in the system's evolutionary history, associated with a recent capture and rapid shredding of a donor star (on a timescale of a few years). HLX-1 should now head into quiescence.

Instead, here we want to consider whether HLX-1 can be a more stable system, capable of supporting high mass transfer for a much longer timescale ({\it e.g.}, $\sim$ 10$^5$ yr rather than $\sim$10 yr). For this, we need to discard the assumption that the outburst recurrence timescale is approximately the same as the orbital period. We will discuss and compare two alternative physical mechanisms that may produce outburst cycles, not based on the disk-instability scenario: one for the case of a small orbital separation, and one for a larger separation (consistent with the optical disk size measured in HLX-1). As we explained in Section 4.3, we will argue that the latter one is the more plausible.

\subsection{A mass transfer instability scenario for the small disk case}

The first scenario assumes a small accretion disk 
  with an outer radius $\la 10^{12}\;\! {\rm cm}$. 
The observed timescale for the luminosity decline corresponds 
  to the time needed for the whole disk to drain, 
  and the outburst is due to an intermittent or variable gas inflow. 
Without the aid of a highly eccentric orbit, 
  an alternative mechanism is needed 
  for the recurrent bursts of mass transfer from the donor star.
Given the strong X-ray luminosity and small orbital separation, 
 irradiation on the donor star is expected to be strong.  
Studies \citep{phillips02} have shown that 
 the pressure exerted by a strong irradiation 
 could press down the surface of the star, 
 causing a temporary loss of contact with the critical Roche surface 
 at the inner Lagrangian point (L1). 
When this happens, the star loses mass instead 
  through the outer Lagrangian point (L2), 
  feeding a CB disk. 
However, moderate irradiation by X-rays could have an opposite effect:  
  irradiative heating may lead to an increase in the donor star's atmospheric scale height,   
  resulting in more mass flowing into the BH's Roche lobe through L1   
  \citep{podsiadlowski91,kovetz88}. 
The importance of irradiation feedback in BH X-ray binaries 
  (especially those with a large accretion disk, which casts a shadow on the star) 
  has been under debate \citep{vilhu94,ritter00,ritter08}, 
  but observations have shown 
  that the process does operate in binary systems 
  containing a white-dwarf accretor \citep{mroz16}. 
Persistent X-ray irradiation could drive  
  cycles of high and low mass transfer rates in low-mass X-ray binaries, 
  with peaks reaching $10^{-5} M_{\odot}\;\!{\rm yr}^{-1}$  
  or higher \citep{harpaz94},  
  but the characteristic duty cycle of $> 10^4\;\!{\rm yr}$   
  implies that this mechanism is irrelevant for the outbursts 
  seen in HLX-1.

With the effects of radiation pressure on the donor star surface in mind, 
  we suggest the following scenario to account 
  for the $\sim 100\;\!{\rm day}$ HLX-1 outburst cycles.  
At the peak of each outburst, 
  the irradiating X-ray pressure becomes high enough 
  to press down the donor star's atmosphere, 
  making the star detach from the critical Roche surface at L1.  
When the disk is not fed, the outburst subside on the viscous timescale. 
After most of the disk is consumed, 
  the system enters a low/hard state, 
  with a luminosity allowing the donor star 
  to regain contact with the critical Roche surface 
  and resume the mass transfer process. 
This sequence of events explain the presence of a CB disk  
  (required to account for the optical luminosity),  
  presumably fed through L2 at the moment that the star becomes detached at L1,   
  in a natural manner. 
This scenario works when  
 the X-ray luminosity $L_{\rm X} \ga 0.8 L_{\rm Edd,2}$ \citep{phillips02} 
 (where $L_{\rm Edd,2}$ is the Eddington luminosity of the star). 
For HLX-1 $L_{\rm X} \approx 10^{42}$ erg s$^{-1}$ during the outburst peak. Such luminosity is $\gg L_{\rm Edd,2} \approx 10^{38}$ erg s$^{-1}$.     
The question is 
  how much accretion luminosity reaches L1 in the presence of shadowing from the outer edge of the disk. 
There are several possibilities for a partial irradiation of L1 even in the presence of a disk shadow \citep{viallet08}. 
For instance, a warped disk may provide the amount of irradiation needed. 
Accretion disks in both the high/soft state and the super-Eddington regime
  could launch strong winds \citep{munoz16,pinto16,ponti12,neilsen09},  
  which can scatter a fraction of the X-ray emission down towards L1. 
Photons can also be scattered towards L1 by an extended hot corona above the inner disk    or near the base of a jet. 
Note that for this to happen, 
  the vertical extent of the X-ray scattering region 
  would have to be $\ga 10^{11}\; \!{\rm cm}$, 
  for a disk with a thickness $H/R \sim$ a few $10^{-2}$. 
From the observed temperatures and luminosities of outer disks 
  in the Galactic X-ray binaries, 
  \cite{viallet08} estimated that 
  about $\sim 10^{-3}$ of the accretion luminosity 
  could be scattered by the disk wind, outflow or corona 
  to irradiate the region near L1. 
The condition $L_1 \ga 0.8 L_{\rm Edd,2}$ is thus easily satisfied, 
  leading to the detachment of the donor star from the critical Roche surface at L1. 
As a comparison, the Galactic source SS\,433 is an example of an X-ray binary where the direct emission is partially occulted or eclipsed, but a residual hard X-ray component 
  (multi-temperature bremsstrahlung and/or Compton scattering emission) 
  is seen even when the disk is eclipsed by the companion star.  
The inferred emission region for this extended component is $\sim 10^{12} -  10^{13}\;\!{\rm cm}$      \citep{marshall12,kubota10,krivosheyev09,cherepashchuk07,kotani96}, 
  which is as large or larger than the size of the accretion disk 
   ($R_{\rm out}  \approx 2 \times 10^{12}$ cm: \citealt{gies02}). 
The observed X-ray luminosity of SS\,433 
  ($L_{\rm X} \approx 10^{36}$ erg s$^{-1}$) 
  is $\sim$10$^{-3}$--10$^{-4}$ times the true luminosity \citep{khabibullin16,fabrika15,medvedev10}, 
  because only a small fraction of X-rays are scattered at all angles 
  by the outflow while most of the radiation is emitted 
  inside the polar-axis funnel 
  \citep{atapin15,medvedev10}. 
If a similar fraction of the X-ray luminosity is scattered 
  by a wind above the disk plane in HLX-1, 
  the pressure exerted by the irradiation flux 
  would satisfy the condition given in \cite{phillips02} 
  and cause an intermittent mass transfer. 
  
In conclusion, we stress that the scenario described in this Section is not the most likely explanation for HLX-1, but we have presented it for completeness, as an example of an at least theoretically possible cycle of variability in a highly luminous, compact system, with a donor star alternatively feeding the inner disk and the CB disk.
  

\subsection{An oscillating wind scenario for the large disk case}



The second scenario considers a system with an accretion disk extending to   $R_{\rm out} \ga 10^{13}\;\! {\rm cm}$. 
The arguments outlined in Section 5.2, about irradiation suppressing mass transfer through L1 and inducing L2 overflow \citep{phillips02}, are also  applicable here. 
The larger binary separation in this case 
  is compensated by a larger cross section for photon interception provided by the larger donor star.  
The larger orbital separation 
  implies that the Roche lobe (Section 3.3.2)  
  can accommodate a more massive donor, 
  and hence permits a higher value for $L_{\rm Edd,2}$; in principle, this makes it harder for the irradiation condition to be satisfied. 
However, in the large-disk scenario, the condition is irrelevant 
  and cannot explain the outburst cycle of HLX-1.  
Here is the reason. 
The amount of material stored 
  in a steady-state standard disk 
  with $R_{\rm out} \sim 10^{13}\;\! {\rm cm}$
  is a few $10^{30}\;\!{\rm g}$ (Equation 4), 
  but only a few $10^{28}\;\! {\rm g}$ are accreted into the BH 
  during each outburst (Section 4.1 and \citealt{yan15}). 
The large outer disk practically is a reservoir 
  that tends to smooth out mass-transfer fluctuations from the donor star 
  on timescales shorter than 
  the fluid viscous timescale at the outer radius (which is $\sim$10$^3$ yr). 
Whether or not the donor star temporarily detaches from L1 
  during each outburst cycle is therefore unimportant, 
  when the amount of mass involved in each outburst duty cycle  
  is only a small fraction of the mass stored in the disk. 
The main feature of the large-disk scenario 
  is that the outburst would terminate 
  when a small part of the disk 
  (presumably, the innermost part) is depleted.  
Thus, the observed alternation of high and low states must be driven 
  by changes in the accretion rate 
 through the disk at $R \la 0.1 R_{\rm out}$ 
 rather than by variations in the mass transfer through L1. 

A comparison of the outburst behaviours of HLX-1 and of the Galactic BH V404 Cyg \citep{kimura16} shows some interesting similarities.
The binary parameters of V404 Cyg 
  suggest that the accretion disk has a radius 
  $R_{\rm out} \approx 9 \times 10^{11}\;\! {\rm cm}$.  
It is believed that during the outburst a massive thermal wind is launched, 
  depleting the accretion disk;  
the wind is strong enough to deplete the mass inflow \citep{munoz16}. 
As a result, the inner disk region 
  (within a radius $R \sim 0.1 R_{\rm out}$) 
  is no longer fed (or is
fed at much lower rates) by the mass inflow from larger radii.  
The outburst subsides on the viscous timescale 
 of the inner part of the disk (inside the wind depletion radius), 
 which is much shorter than the viscous timescale of the whole disk. 
An estimate of only $\approx 0.1$ percent of the disk mass is accreted 
 during an outburst \citep{munoz16}. 
The optical continuum is mostly reprocessed emission from the irradiated surface of the full accretion disk. 
A consequence of the massive disk outflow 
  is the formation of a large H$\alpha$-emitting nebula. 
These features of V404 Cyg 
 are qualitatively similar to what we observe in HLX-1. 

Another property that V404 Cyg and HLX-1 have in common 
  is the relatively high luminosity in the inter-outburst periods. 
At $L_{\rm X} \approx 10^{33}$ erg s$^{-1}$, 
  V404 Cyg is two orders of magnitude more luminous than 
  typical BH X-ray binaries in quiescence \citep{munoz16,rana16,armas14}. 
Even so, the ratio of outburst to inter-outburst luminosities is still as high as $\sim$10$^5$ in V404 Cyg, while it is only $\approx$30 in HLX-1.
The anomalously high luminosity in the inter-outburst periods 
  can be explained by the continuing presence of low-level accretion 
  from the already detached outer disk, 
  which is not completely depleted after the outburst 
  and is working as a reservoir \citep{munoz16}. 
In V404 Cyg, the double-peaked H$\alpha$ emission line 
  in the inter-outburst period
  is strong evidence that the outer disk survives 
  through quiescence \citep{munoz16}. 
In this scenario, the different ratio of outburst to inter-outburst luminosities suggests that the wind-driven disruption of the inner accretion flow is not as strong in HLX-1 as it is in V404 Cyg.


If the disk wind is strong enough to play a major role 
  in regulating the outbursts of HLX-1,    
  a damped disk-wind instability \citep{begelman83,shields86} would be inevitable. 
This instability occurs 
  when accretion, 
  at a rate $\dot{M}_{\rm a}$ in the X-ray emitting region of the disk, 
  triggers a massive wind in a larger annulus of the disk 
  (at $R \equiv R_{\rm w} < R_{\rm out}$), 
  with an outflow rate $\dot{M}_{\rm w}(t)  \equiv C \dot{M}_{\rm a}(t)$, 
  with $C \ga$ a few. 
For this to happen, 
  the wind has to be directly induced by irradiation. 
The original model put forward by \cite{begelman83} and \cite{shields86}   considered a thermal (Compton) wind,  
  but the mechanism would also work with a radiatively driven wind. 
In a steady state, 
  $\dot{M}_{\rm in} = \dot{M}_{\rm a} + \dot{M}_{\rm w} = (1+C)\, \dot{M}_{\rm a}$, 
   where $\dot{M}_{\rm in}$ is the accretion rate in the outer region of the disk 
   beyond the wind-dominated region. 
A small positive perturbation $\delta \dot{M}_{\rm a}$ 
  will cause a prompt increase 
  $\delta \dot{M}_{\rm w} = C \times \delta \dot{M}_{\rm a}$ of the outflow rate. 
As a consequence, $\dot{M}_{\rm in} < \dot{M}_{\rm a} + \dot{M}_{\rm w}$, 
  and the disk is depleted inside $R_{\rm w}$. 
While the wind immediately reacts to the X-ray luminosity increase, 
  it takes a viscous diffusion timescale for $\dot{M}_{\rm a}$ 
  to feel the effect of the depleted inflow arriving from $R_{\rm w}$ 
  and to re-adjust to a lower value. 
When that happens, the wind immediately decreases, 
  so that $\dot{M}_{\rm in} > \dot{M}_{\rm a} + \dot{M}_{\rm w}$ 
  and the inner disk builds up again.
It is the difference in the response timescales of the radiation-induced wind  
  and of the inner-disk accretion that creates and sustains the oscillations. If $C\ga 10$, oscillations are persistent;   
for $1 < C \la 10$, they are damped after a few cycles \citep{shields86}. 
The amplitude of the oscillations is
   $\dot{M}_{\rm a,max}/\dot{M}_{\rm a,min} \approx 10$ \citep{shields86}.
Using the standard Shakura-Sunyaev viscosity parameterization,  
   \cite{shields86} found an oscillation period $P_{\rm osc} \sim 0.1$--0.4 $\tau_{\rm w}$,  
   where $\tau_{\rm w}$ is the viscous timescale in the wind region. 
The period increases when the wind efficiency parameter $C$ decreases,  but is only a weak function of $C$.  

We suggest that this model is applicable to HLX-1, 
 which is expected to have strong disk winds in outburst. 
As a comparison, Galactic low-mass X-ray binaries in the high state,  
  with $L_{\rm X} \sim 0.1 L_{\rm Edd}$, 
can have a mass outflow rate in a disk wind  
  $\sim$1--20 times the accretion rate \citep{diaztrigo16}, 
  which is strong enough to allow the wind instability 
  described above to develop. However, we need to estimate whether the wind instability predicts the correct outburst timescale.

For a thermal wind, 
  the launching radius is $R_{\rm w} \sim 0.1 R_{\rm IC}$ 
  \citep[see][]{shields86}, 
  where $R_{\rm IC}$ is the Compton radius, 
  where the isothermal sound speed of the gas 
  equals the escape velocity from the disk. 
  For an irradiating source spectrum peaking at energies 
  $\approx 1 \;\! {\rm keV} (\approx 10^7\;\! {\rm K})$, 
  the Compton radius $R_{\rm IC} \sim 10^{11} (M/M_{\odot})\;\!{\rm cm}$. Therefore, in HLX-1, thermal winds could only be launched if the disk extended to $\sim 10^{14}$ cm, which does not appear to be the case.
However, when the effect of radiation pressure is also included,
  strong winds can be launched 
  from smaller radii: in particular, \cite{proga02} showed that UV lines can drive powerful winds from regions of the disk that have a surface temperature $\la 50,000$ K.
Based on the results of our broad-band modelling, we have estimated (Section 4.2) an effective temperature $T_{\rm eff} \approx 20,000$ K at $R \approx R_{\rm out} \approx (2 \times 10^{13})/\sqrt{\cos \theta}$ cm in the high state. In the irradiation-dominated part of the disk, $T_{\rm eff} \propto R^{-1/2}$ \citep{dubus99}. Thus, we expect a temperature $T_{\rm eff} \approx 50,000$ K at $R \approx R_{\rm w} \approx 3 \times 10^{12}$ cm $\approx 3 \times 10^{-3} R_{\rm IC}$. (As a comparison, for X-ray binaries $R_{\rm w} \sim 5 \times 10^{-2} R_{\rm IC}$, and for AGN, $R_{\rm w} \sim 10^{-4} R_{\rm IC}$: \citealt{proga02}).
 
Assuming a Shakura-Sunyaev disk \citep{ss73,frank02} with Kramers opacity and $\dot{M} \sim 10^{21}$--$10^{22}\;\!{\rm g~s}^{-1}$ (spanning the range of plausible accretion rates for HLX-1 in the low and high state), 
  we obtain a viscous timescale $\tau \sim $ a few years at $R \sim$ a few  $10^{12}$ cm. (This can be verified simply by recalling that $\tau \sim \left[\alpha \, \Omega \, (H/R)^2\right]^{-1} \approx 6$ yr for $(H/R) \approx 0.01$, $\alpha \approx 0.3$, $M \approx 2 \times 10^4 M_{\odot}$.) Therefore, the wind oscillation period $P_{\rm osc}$ can be as short as $\approx$1 year \citep{shields86}. 
The observed outburst duty cycle in HLX-1 
  is consistent with this scenario. 
The increase in the outburst recurrence period, 
  accompanied by a reduced fluence in each cycle \citep{yan15}, 
  is also consistent with damping of the oscillations, 
  as the wind multiplication factor decreases. 
  
Line-driven winds are effective when the UV luminosity $L_{\rm UV} > L_{\rm Edd}/M_{\rm max}$ \citep{murray95,proga98,proga02}, where $M_{\rm max}$ is the limit of the force multiplier, which can be as high as $\approx$2000 for low-ionization gas. In the case of HLX-1, for the 2010 high/soft state spectrum we infer an unabsorbed luminosity in the 13.6--100 eV band $L_{\rm UV} \approx 0.15 L_{\rm X} \sim$ a few $10^{-2} L_{\rm Edd}$. Therefore, we can at least say that there are enough UV photons available in the system to drive a strong wind. However, the amount of mass carried in the wind depends on the ionization parameter of the gas, and hence on the amount of shielding from direct X-ray irradiation, which would over-ionize the wind and reduce the effectiveness of line driving. We do not have any empirical constraints on the wind parameters for HLX-1, perhaps because our line-of-sight is almost face-on, not intersecting the wind. Thus, a more detailed modelling of the energetics of the wind is beyond the scope of this work. However, we do note that optical spectra of HLX-1 have shown an H$\alpha$ emission line with a luminosity of a few $10^{37}$ erg s$^{-1}$ and full width at half maximum of $\approx$400 km s$^{-1}$ \citep{wiersema10,soria13b}. The line is too narrow to be directly associated with a disk wind, but it suggests at least the presence of diffuse gas around the system.

A wind-driven sinusoidal oscillation in $\dot{M}_{\rm a}$ 
 does not necessarily result in a sinusoidal oscillation in $L_{\rm X}$,  because accreting BHs have distinct accretion states, and hysteresis 
 in their cycle of states. 
If $\dot{M}_{\rm a}$ exceeds a few percent of $\dot{M}_{\rm Edd}$ 
  at the peak of the oscillation, 
  the system will switch to the high/soft state, before returning to the low/hard state during the low part of the oscillation. 
The fact that the time spent in the high/soft state was getting shorter from cycle to cycle \citep{yan15}, and no new outbursts have been detected for a couple of years (since 2015 January), 
  may both be a consequence of the damping of the oscillation. 
We suggest that the system is now settled in the low/hard state, 
  where the long-term-average accretion rate 
  is below the state-transition threshold 
  of a few percent of the Eddington accretion rate,  
  and the damped oscillation in $\dot{M}_{\rm a}$ 
  does not reach the threshold for a hard-to-soft state transition any more. 
  For example, we suggest a scenario in which the long-term-average accretion rate is $\approx 0.05 \dot{M}_{\rm Edd} \approx 10^{21}$ g s$^{-1}$ $\approx 2 \times 10^{-5} M_{\odot}$ yr$^{-1}$ (for an efficiency $\eta \approx 0.1$), just below the threshold for the collapse of the hot, radiatively inefficient corona and the resulting hard-to-soft transition. By analogy with Galactic X-ray transients, the radiative luminosity in this state is expected to be $\approx 0.02 L_{\rm Edd} \approx 4 \times 10^{40}$ erg s$^{-1}$ (consistent with the luminosity inferred from {\it Swift} since the end of the last outburst), with an additional similar contribution to the accretion power in the form of kinetic power of a jet\footnote{Here, we have assumed that the low/hard state is less radiatively efficient than the high/soft state: $\eta_{\rm lh} \approx 0.05$ at $\dot{M} \approx 0.05 \dot{M}_{\rm Edd}$, with a similar amount of power going into the jet.}.  A positive perturbation in the accretion rate would push the system over the threshold for the high/soft state at $L \sim 0.1 L_{\rm Edd}$ (briefly peaking at $L \approx 0.3 L_{\rm Edd}$ for a few days after the transition). The subsequent wind-driven oscillation of the accretion rate below $\approx 0.02 \dot{M}_{\rm Edd}$ would bring the system back to the hard state. 
  
  An accretion rate $\dot{M} \approx 10^{21}$ g s$^{-1}$ is very high but comparable to the accretion rates proposed for the Galactic source SS\,433 and for the most powerful ULXs in nearby galaxies \citep{feng11,sutton13,fabrika15,bachetti16}, which are thought to be super-Eddington, stellar-mass accretors. For a 10-$M_{\odot}$ BH, an accretion rate $\dot{M} \approx 10^{21}$ g s$^{-1}$ corresponds to $\dot{m} \equiv \dot{M}/\dot{M}_{\rm Edd} \approx 100$ and a luminosity $L \approx (1 + 0.6 \ln \dot{m}) \approx 4 L_{\rm Edd} \approx 5 \times 10^{39}$ erg s$^{-1}$ \citep{poutanen07}. 
In HLX-1, for a BH mass $M \approx 2 \times 10^4 M_{\odot}$ and a binary separation $a \approx 4 \times 10^{13}$ cm (Section 3.3.2), the binary period is $P \approx 10$ d. 
A donor star of mass $M_2 \approx 15 M_{\odot}$ 
  is able to fill its Roche lobe at a radius $R_2 \approx 25 R_{\odot}$ 
  (typical of a blue supergiant). 
In summary, this scenario does not require that we are {\it observing} the system during a one-off, short-duration event, such as the tidal stripping of a recently-captured star over a few orbital timescales. The average mass transfer rates can instead be sustained for $\sim$10$^5$ yr (or $\sim$10$^4$ for a red-giant donor). It still does require peculiar {\it formation} conditions, such as a donor star in a close orbit around an IMBH, filling its Roche lobe as it evolves off the main sequence. The likelihood of such event is hard to estimate with the constraints at hand, and not knowing what fraction of compact stellar systems contains an IMBH in their core (\citealt{caputo17} estimate the existence of $\sim$10$^6$ IMBHs with masses $\sim$10$^4 M_{\odot}$ within 100 Mpc). On the other hand, from the observational point of view, HLX-1 is a unique system within at least 100 Mpc, and we should not discount one-off events. If our oscillation scenario is correct, we predict that the X-ray luminosity of HLX-1 over the next few years will remain at $\ga$10$^{40}$ erg s$^{-1}$; instead, if short-duration tidal stripping scenarios are correct, HLX-1 should soon fade into quiescence.

\section{Conclusions}


We compared the optical/UV brightness of HLX-1 during three sets of {\it HST} observations taken at different times during the outburst cycle. We fitted the optical/UV data together with X-ray data taken at similar phases, to quantify how the optical/UV luminosity is a function of the X-ray flux. We showed that the optical/UV emission is well modelled with at least two components: a blue component, strongly dependent on the irradiating X-ray flux, and a constant red component. The latter is consistent with an underlying old stellar population, perhaps a globular cluster or the nucleus of a satellite dwarf. The bluer optical emission is dominated by an irradiated disk with an outer radius $R_{\rm out} \sqrt{\cos \theta} \approx 2 \times 10^{13}$ cm. The high level of irradiation of the outer accretion disk disfavours models based on strongly beamed X-ray emission (as proposed by \citealt{king14} if HLX-1 were a highly super-Eddington stellar-mass accretor), or at the very least it would require a peculiar wind structure in the polar funnel, in order to scatter a significant amount of photons onto the outer disk.
The data are also consistent with an additional blue component coming from a young stellar population. The residual optical emission in the 2013 observations (when both the X-ray source and the optical counterpart were at their faintest level) clearly places an upper limit to the (constant) contribution of the young stellar population: we find that a single-population young stellar cluster must have a mass $\la 10^4 M_{\odot}$ to avoid exceeding the blue luminosity in the 2013 data.

From our analysis and comparison of {\it XMM-Newton} and {\it Swift} lightcurves and spectra, we argued that the 2012 November observations were fortuitously taken right at the time of the hard-to-soft transition at the end of an outburst cycle. As a result, based on the analogy with similar transitions in Galactic X-ray transients, we argued that the X-ray luminosity at that point was $\approx (2 \pm 1)\%$ of the Eddington luminosity. Combining this constraint with the independent constraint on the inner disk size from spectral fitting ($R_{\rm in} \sqrt{\cos \theta} \approx 50,000$ km), we estimated a BH mass $\approx (2^{+2}_{-1}) \times 10^4 M_{\odot}$, with a spin parameter $a/M \ga 0.9$ for a moderately face-on viewing angle. This is in agreement with the results of \cite{davis11} based only on detailed spectral modelling of the disk spectrum in the high state. For these system parameters, the peak observed luminosity in the high/soft state is $L_{\rm X} \approx 0.2$--$0.4 L_{\rm Edd}$ (similar to most Galactic BH transients), and the luminosity in the low/hard state is $L_{\rm X} \approx 0.01 L_{\rm Edd}$. With a different assumption of the geometry of emission in the intermediate state (isotropic rather than disk-like), we obtain a slightly different BH mass, $M \approx (3^{+3}_{-1}) \times 10^4 M_{\odot}$, but the qualitative result is the same: the system oscillates between sub-Eddington, canonical high/soft and low/hard states.

We then discussed the size of the accretion disk, based on two apparently contradicting pieces of information. The short timescale of the X-ray rise, decay and recurrence time points to a small disk, with $R_{\rm out} \la 10^{12}$ cm, and more likely $\sim$ a few $\times 10^{11}$ cm. Instead, the reprocessed UV/blue emission requires a surface with a radius $R_{\rm out} \ga 2 \times 10^{13}$ cm. To reconcile the two observations, we presented and compared two alternative scenarios, both based on a two-zone structure for the accretion flow: an inner region ($\la 10^{12}$ cm) that drives the X-ray outburst, and an outer region ($\ga 2 \times 10^{13}$ cm) that produces most of the optical/UV luminosity. In the first scenario, the inner region is the accretion disk inside the BH Roche lobe, and the outer region is a CB disk, with a gap in between due to the orbit of the donor star. The outburst may be triggered by a mass transfer instability due to the high irradiation seen by the donor star. The star would alternately feed the accretion disk (through the Lagrangian point L1) or the CB disk (through L2) depending on the irradiation condition. In the second scenario, both X-ray and optical/UV emission come from a large accretion disk. However, only the inner region of this inflow produces the X-ray outburst cycles on a short timescale; the outer disk acts as a long-term reservoir of matter. In this scenario, the mass accretion flow towards the inner region oscillates due to the feedback effect (wind instability) of radiatively-driven outflows, which deplete the middle region of the disk and regulate the flow that goes past a radius $\sim 10^{12}$ cm.
The latter scenario is similar to the current interpretation of the outburst in the Galactic BH V404 Cyg. 

Based on the available optical/UV and X-ray luminosity alone, it is not possible directly to rule out either scenario. The spectral energy distribution expected from an irradiated CB disk is similar to the one expected from an irradiated outer disk of similar size (in both cases, $T \propto R^{-1/2}$). The only difference would be the presence and size of a gap at $R \approx 10^{12}$ cm in the case of a CB disk, due to the orbit of the donor star. The gap would affect disk annuli with a temperature $T_{\rm eff} \approx 10^5$ K, and would produce a small decrease in the far-UV flux, a spectral band that is not directly observable. 
However, the two scenarios have very different implications for the size and type of donor star and for the binary period. A binary separation of $\approx$1--$2 \times 10^{12}$ cm (small-disk scenario) implies a binary period of only $\approx$2--3 hr and a very small Roche lobe for the donor star, consistent only with the size and density of an M-dwarf close to its tidal disruption limit. A large-disk scenario implies a more plausible binary period $\sim$10 d, and the secondary Roche lobe can accommodate a blue supergiant or other types of post-main-sequence stars, able to provide a higher mass transfer rate over a longer timescale. If the donor star is a blue supergiant, this would confirm the presence of a small population of young stars around the IMBH, but the data do not rule out an older donor (ascending the red giant branch).
Based on these arguments, we conclude that the large-disk scenario with wind-driven oscillations in the inner region is more consistent with the data and is the most plausible explanation for HLX-1. 

More generally, the wind-driven oscillation scenario provides a relatively unexplored mechanism for the generation of super-orbital X-ray variability (in the case of HLX-1, with a super-orbital period $\approx$40 times longer than the binary period), in alternative to the inner-disk/jet precession scenario. We speculate that the same mechanism may explain other examples of super-orbital periodicities in luminous X-ray binaries \citep{farrell06,farrell09b} and ULXs ({\it e.g.}, the 115-d periodicity in NGC\,5408 X-1: \citealt{strohmayer09,foster10,grise13}).


According to our adopted scenario, the system is now in the low/hard state at $L_{\rm X} \sim 10^{-2} L_{\rm Edd}$, only a factor of a few below the accretion rate and luminosity at which we expect a transition to the high/soft state. We suggest that the oscillations in the accretion rate have damped to the point where even at its peak, $\dot{M}$ is no longer high enough to trigger the transition. We do not have empirical evidence to test the origin of the perturbation that caused the damped oscillatory behaviour over the last decade--it could be related to the evolution of the donor star, or to a super-orbital period, or just stochastic variability. In engineering terms, the radiatively-driven wind acts as a proportional controller on $\dot{M}_{\rm a}(t)$, with a lag time of order of the viscous timescale. The setpoint is a luminosity slightly below the threshold for the hard-to-soft transition. If the gain of the proportional controller is sufficiently high, the lag time inevitably causes oscillations. In earlier cycles, when the oscillating error was high enough, the system underwent a state transition.

\section*{Acknowledgements}

We thank Chris Copperwheat, Fabien Gris\'{e}, Jifeng Liu, Michela Mapelli, Claudia Maraston, James Miller-Jones, Manfred Pakull, Megumi Shidatsu, Vlad Tudor, Yoshihiro Ueda and the anonymous referee for useful suggestions and discussions. RS acknowledges support from a Curtin University Senior Research Fellowship. He is also grateful for support, discussions and hospitality at the Strasbourg Observatory and at the Aspen Center for Physics during part of this work. 
LZ acknowledges funding from the ASI-INAF contract {\it NuSTAR} I/037/12/0. EvdH acknowledges support from NWO (grants Number 643.200.503, 639.073.803 and 614.061.608) and NOVA as well as supervision from S.~Portegies Zwart. This paper benefitted from discussions at the 2016 International Space Science Institute workshop ``The extreme physics of Eddington and super-Eddington accretion on to Black Holes'' in Bern, Switzerland (team PIs: Diego Altamirano \& Omer Blaes). 
The quasi-simultaneous 2012 and 2013 {\it XMM-Newton} and {\it HST} data were obtained through the joint {\it XMM-Newton}/{\it HST} time allocation program.
This research has made use of the NASA/IPAC Extragalactic Database (NED) which is operated by the Jet Propulsion Laboratory, California Institute of Technology, under contract with the National Aeronautics and Space Administration. We also used Leo C.~Stein's ``Kerr ISCO Calculator'' online.












\bsp	
\label{lastpage}
\end{document}